\documentclass[pra,twocolumn,groupedaddress,superscriptaddress,amsmath,amssymb,a4paper]{revtex4} 
\usepackage{geometry}
\usepackage{graphicx}
\usepackage{verbatim}
\usepackage{float}
\usepackage{graphics}
\usepackage{mathrsfs}
\usepackage{amssymb}
\usepackage{amsmath}
\usepackage{amsthm}
\usepackage{bm}
\usepackage{hyperref}
\usepackage{braket}
\usepackage{mathtools}
\usepackage{pst-node}
\usepackage{amsbsy}
\usepackage{tikz}
\usetikzlibrary{calc}

\usepackage[nodisplayskipstretch]{setspace} 

\newcommand{\tikzmark}[1]{\tikz[overlay,remember picture] \node (#1) {};}
\newcommand{\DrawBox}[3][]{%
    \tikz[overlay,remember picture]{%
        \coordinate (TopLeft)     at ($(#2)+(-0.6em,0.9em)$);
        \coordinate (BottomRight) at ($(#3)+(0.6em,-0.3em)$);
        \path (TopLeft); \pgfgetlastxy{\XCoord}{\IgnoreCoord};
        \path (BottomRight); \pgfgetlastxy{\IgnoreCoord}{\YCoord};
        \coordinate (LabelPoint) at ($(\XCoord,\YCoord)!0.9!(BottomRight)$);
        \draw [red,#1] (TopLeft) rectangle (BottomRight);
    }
}

\def\VR{\kern-\arraycolsep\strut\vrule &\kern-\arraycolsep}

\newcommand{\rn}[2]{
    \tikz[remember picture,baseline=(#1.base)]\node [inner sep=0] (#1) {$#2$};%
}

\newcommand{\ave}[1]{\langle #1 \rangle}
\newcommand{\ip}[2]{{\langle #1|}{ #2 \rangle }}

\addcontentsline{toc}{chapter}{References}

\begin{document}

\title{Effect of incoherent pump on two-mode entanglement in optical parametric generation}

\author{S. V. Vintskevich}
\affiliation{Moscow Institute of Physics and Technology,
Institutskii Per. 9, Dolgoprudny, Moscow Region 141700, Russia}

\author{D. A. Grigoriev}
\affiliation{Moscow Institute of Physics and Technology,
Institutskii Per. 9, Dolgoprudny, Moscow Region 141700, Russia}

\author{S. N. Filippov}
\affiliation{Moscow Institute of Physics and Technology, Institutskii Per. 9, Dolgoprudny, Moscow Region 141700, Russia}
\affiliation{Valiev Institute of Physics and Technology of Russian Academy of Sciences, Nakhimovskii Pr. 34, Moscow 117218, Russia}
\affiliation{Steklov Mathematical Institute of Russian Academy of Sciences, Gubkina St. 8, Moscow 119991, Russia}

\begin{abstract}
Pumping a nonlinear crystal by an intense radiation results in the optical parametric generation of photons in two modes (the signal and the idler). The quantized electromagnetic field in these modes is described by a continuous-variable quantum state, which is entangled if the pump is a coherent state produced by a laser. The signal and the idler modes remain populated by photons even if the pump becomes incoherent (dephased by a medium, superposed with a thermal state, or produced by an alternative source such as the superluminescent diode). However, the incoherent pump does effect the entanglement and purity of the signal and the idler modes, which is of vital importance for the quantum information applications and the interferometry. Here we develop an approach to infer the signal-idler entanglement and purity for a general quantum incoherent pump with the given Glauber-Sudarshan function. We show that the signal-idler entanglement is extremely sensitive to the phase distribution of the pump and illustrate our findings by physically relevant examples of the incoherent pump: the noisy coherent state, slightly dephased and phase-averaged coherent states, thermal state, and states modulated by the Kerr medium. The effect of incoherent pump on the combined quadratures is discussed as well.
\end{abstract}


\maketitle

\section{Introduction}

Optical parametric generation (OPG) is a basis of modern quantum optics with numerous applications in quantum information and fundamental tests of quantum mechanics (see the reviews \cite{Braunstein,Illuminati,KOK,FedorovKulik,Lvovsky,PANJW} and references therein). In the process of the OPG, a pump light beam propagates through a nonlinear medium and an interaction between the pump mode and vacuum fluctuations of the electromagnetic field occurs. A quantum description of such a process was suggested in the papers~\cite{11,12}. Zel'dovich and Klyshko have noticed that the pump photons with frequency $\omega_p$ break up into pairs of scattered quanta  with lower frequencies $\omega_i$ and $\omega_s$ in accordance with energy conservation law $\omega_{p}=\omega_{i}+\omega_{s}$~\cite{Zeldovich}. Down-converted photons are usually called idler ($i$) photon and signal ($s$) photon, also referred to as  biphotons in the spontaneous parametric down-conversion process~\cite{Klyshko2018}. Burnham and Weinberg experimentally confirmed simultaneity in the production of photon pairs~\cite{Burnham}. The pump wave vector $\vec{k}_p$ in the crystal satisfies the phase matching condition, which is a momentum conservation law for down converted photons, $\vec{k}_{p} \approx \vec{k}_{i}+\vec{k}_{s}$.  The approximate equality is due to the finite size of a real crystal and a finite spectral width of the pump beam. The combination of the momentum and energy conservation laws determines the angular and frequency distribution of the down-converted photons, the entanglement properties, and the quantum interference characteristics~\cite{PANJW,Howell,PRA_MVF,Fedorov2018,V2019}.

The use of avalanche photodiodes as detectors enables one to effectively probe the single-photon subspace of the idler and signal modes described by the state
\begin{equation}\label{PSI_DCP2}
\ket{\psi} = \sum_{\alpha,\beta = H,V}\int d\vec{k}_{i}d\vec{k}_{s}F\left(\vec{k}_{i},\alpha;\vec{k}_{s},\beta\right) a_{\vec{k}_{i},\alpha}^{\dagger }a_{\vec{k}_{s},\beta}^{\dagger}\ket{{\rm vac}},
\end{equation}

\noindent where $H$ and $V$ denote the horizontal and vertical polarizations, respectively, $(\vec{k},\alpha)$ is the mode of electromagnetic radiation, $a_{\vec{k},\alpha}^{\dag}$ is a creation operator for photons in the mode $(\vec{k},\alpha)$, and the function $F\left(\vec{k}_{i},\alpha;\vec{k}_{s},\beta\right)$ is a biphoton's wavefunction. Experiments with the polarization degrees of freedom~\cite{CASTELLI} operate with the polarization density operator
\begin{eqnarray} \label{dv}
\varrho_{\alpha\beta,\alpha'\beta'} &=& \int_{\Omega} d\vec{k}_i d\vec{k}_s  F\left(\vec{k}_{i},\alpha;\vec{k}_{s},\beta\right) F^{\ast}\left(\vec{k}_{i},\alpha';\vec{k}_{s},\beta'\right) \nonumber\\ 
&& \qquad \quad \times \ket{\alpha,\beta} \bra{\alpha', \beta'},
\end{eqnarray}

\noindent where $\Omega$ is a region of vectors pointing to the polarization detectors. Eq.~\eqref{dv} is a typical example of the \emph{discrete-variable} quantum state of two qubits. 

The important feature of Eq.~\eqref{PSI_DCP2} is that it omits the vacuum contribution because the avalanche photodiodes are only sensitive to the presence of photons. Hence, the states \eqref{PSI_DCP2} are conditional and cannot be created on demand. Eq.~\eqref{PSI_DCP2}  also neglects the contribution of higher number of photons in the idler and signal modes because the probability to observe multiple photons in the signal and the idler is less than that for single photons. However, the use of homodyne detectors~\cite{Smithey,Leonhardt,Lvovsky} and photon-number resolving measurements~\cite{Achilles,Kardynal2008,Divochiy} allows one to probe the contribution of vacuum and multiple photons in the signal and the idler. The corresponding quantum state belongs to the Fock space and reads
\begin{eqnarray}\label{PSI_DCPMul}
\ket{\psi} &=& \sum_{n=0}^{\infty}\sum_{\alpha,\beta = H,V}\int d\vec{k}_{i}d\vec{k}_{s} F_{n}\left(\vec{k}_{i},\alpha;\vec{k}_{s},\beta\right) \nonumber\\
&& \qquad \quad \times (a_{\vec{k}_{i},\alpha}^{\dagger})^n (a_{\vec{k}_{s},\beta}^{\dagger})^n \ket{{\rm vac}}.
\end{eqnarray}

\begin{figure}
\includegraphics[width=8.5cm]{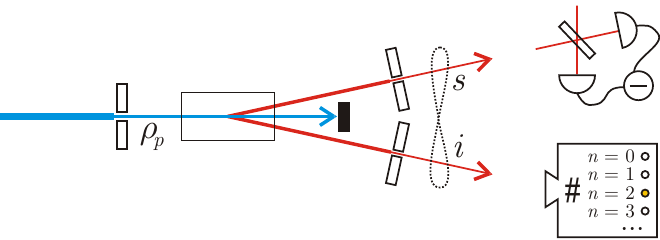}
\caption{Schematic of the experiment with mode filtering to study the effect of incoherent pump $\varrho_p$ on the entanglement of continuous-variable two-mode states (signal mode `$s$' and the idler mode `$i$'). States can be probed by the optical homodyne tomography (top right) and the photon-number resolving detectors (bottom right).}
\label{figure-1}
\end{figure}

Narrow spatial filtering and polarization filtering allows one to \emph{fix} the idler and the signal modes experimentally, see Fig.~\ref{figure-1}. Defining the Fock states $\frac{1}{n!} (a_{\vec{k}_{i},\alpha}^{\dagger})^n (a_{\vec{k}_{s},\beta}^{\dagger})^n \ket{{\rm vac}} = \ket{n_i n_s}$ and the coefficients $c_n = n! F_{n}\left(\vec{k}_{i},\alpha;\vec{k}_{s},\beta\right)$, we get the two-mode state
\begin{equation} \label{PSI_DCP_CV}
\ket{\psi} = \sum_{n=0}^{\infty} c_{n}\ket{n_{i}n_{s}}.
\end{equation}

Equation ~\eqref{PSI_DCP_CV} defines a two-mode \emph{continuous-variable} quantum state because the electromagnetic field amplitude in each mode [eigenvalue $x$ of the operator $X = (a + a^{\dag})/ \sqrt{2}$] belongs to the continuous interval $(-\infty,+\infty)$. In other words, the state~\eqref{PSI_DCP_CV} can be described by a continuous wavefunction $\psi(x,y) = \ip{x,y}{\psi} = \sum_{n=0}^{\infty} c_n \psi_n(x) \psi_n(y)$, where $X\ket{x} = x\ket{x}$, $Y\ket{y} = y\ket{y}$, and $\ip{x}{n} = \psi_n(x) = \pi^{-1/4} (2^n n!)^{-1/2} H_n(x) \exp(-x^2/2)$, with $H_n(x)$ being the the Hermite polynomial of degree $n$. In this paper, we study pure and mixed continuous-variable states of the signal and the idler. 

A paradigmatic example of a pure continuous-variable state is the two-mode squeezed vacuum (twin-beam) state with $c_n = (\cosh r)^{-1} (\tanh r)^n$, where $r$ is the squeezing parameter. Such a state is created in the ideal parametric down conversion~\cite{Barnett1985,Gerry2005}. The experimental implementation of twin-beam states with $10 \log_{10} e^{2r} = 6.2 \div 8.4$ and $\ave{n_i} = \ave{n_s} = \sinh^2 r = 0.6 \div 1.3$ via a non-degenerate optical parametric amplifier is reviewed in Refs.~\cite{wang-2010,yan-2012,zhou-2015,yu-2016}. The two-mode squeezed vacuum is entangled for all $r>0$, i.e. $\psi(x,y) \neq \varphi(x) \chi(y)$. The greater the squeezing parameter, the more entangled the two-mode squeezed vacuum is. Entanglement of such type is useful for many quantum information protocols~\cite{Braunstein, Adesso, Weedbrook,Seifoory} and interferometry~\cite{Anisimov2010}. In particular, the two-mode entanglement is necessary for quantum teleportation of continuous-variable quantum states~\cite{Braunstein1998,Furusawa2007,Chirkin2009,Adesso2017} and entanglement distillation~\cite{kurochkin-2014}. 

The usefulness of entangled continuous-variable states in real experiments is limited by various reasons:  (i) losses and additional classical noise (e.g., in atmospheric turbulence)~\cite{Semenov2009,URen2009,Semenov2010,Martinelli2010,FZ2014,Semenov2016,Bohmann2016,Kunz2018}; (ii) the unavoidable noise in deterministic linear amplifiers~\cite{Caves1982} and general quantum channels~\cite{Holevo2007,filippov2019}; (iii) noise in the preparation of entangled states~\cite{Degiovanni2007,Degiovanni2009,Cesar2009,Arkhipov2015}. General noise results in the purity degradation (the state becomes mixed) and entanglement degradation (the state becomes separable if the noise is strong enough~\cite{Sabapathy2011,FZ2014}). To fight with reasons (i) and (ii), one resorts to the most robust entangled states~\cite{FZ2014,FFK2018} or interventions in the noisy dynamics~\cite{bullock-2018}. Reason (iii) significantly affects the performance of entanglement-enabled protocols because any such protocol relies on a properly prepared (pure) entangled state~\cite{Braunstein, Adesso, Weedbrook,Seifoory}. It is the goal of this paper to study the quality of entangled two-mode continuous-variable states prepared in the OPG with imperfect pump. 

A real pump beam has two types of imperfections: (i) the pump beam is not an infinite plane wave but rather has a specific shape with temporal and spatial coherence properties; (ii) even if a particular plane wave mode is filtered from the pump beam, the quantum state of that mode is not a perfect coherent state $\ket{\alpha}$ in the phase space. The influence of effect (i) on the biphoton density matrix is studied in Refs.~\cite{Giese_2018,CLF1,JHA,CLF2,CLF3,CLF4,CHEKHOVA,LIT1,LIT2,LIT3,CASTELLI}. Instead, we study the influence of effect (ii) on the quality of two-mode continuous variable quantum states as it imposes fundamental limitations on the use of mixed pump states. We focus on a single-mode pump state and  its quantum properties in the phase space. The idea to consider the pump as a general mixed state is motivated by the recent research of the OPG pumped by a light--emitting diode~\cite{LIT1,LIT2,LIT3}. As we show in this paper, the use of the bright thermal light as a pump produces correlated photons in the signal and idler modes, $\varrho = \sum_n p_n \ket{n_i n_s} \bra{n_i n_s}$, however, this form of correlations is classical and there is no entanglement between the modes. 

The effect of the squeezed pump on the properties of the signal and idler modes is studied in Refs.~\cite{HILLERY,Hillery1995,Cohen1995,Hillery2009} and the effect of the pump depletion is analyzed in the degenerate OPG in Refs.~\cite{Cohen1995,QP8}. In this paper, we go beyond the coherent and squeezed pump models and consider a general pump state $\varrho_p = \int P(\alpha) \ket{\alpha}\bra{\alpha} d^2 \alpha$, where $P(\alpha)$ is the Glauber-Sudarshan function~\cite{GLAUBER,Sudarshan,VOGEL}. We show that the structure of the pump state in the phase space significantly affects the purity and entanglement of the signal and idler modes. We illustrate our findings by practically relevant examples of a phase-smeared coherent state, a convolution of thermal and coherent states, and a coherent state affected by the Kerr medium. Our analysis is relevant to all the situations, where the coherence of the pump matters, e.g., in nonlinear interferometers~\cite{CHEKHOVA,FLOREZ}.

The paper is organized as follows.

In Section~\ref{section-density}, we derive the two-mode density operator $\varrho_{is}$ of signal and idler for a general mixed pump within two approaches: the perturbation theory and the general parametric approximation. In Section~\ref{section-purity}, we discuss the purity of $\varrho_{is}$ and the entanglement quantifier (negativity) for $\varrho_{is}$. In Section~\ref{section-thermal}, we analyze the purity dynamics and the entanglement dynamics for the two-mode states in the OPG pumped by the intense thermal state. In section~\ref{section-noisy-coherent}, we study the effect of noisy coherent pump on the quality of the idler-signal entanglement. In Section~\ref{section-phase}, the phase-smeared coherent pump is considered. In Section~\ref{section-general}, we propose an approach to deal with a general mixed pump by applying the Schmidt decomposition to the Glauber-Sudarshan function $P(|\alpha|e^{i\theta})$ of the pump state with respect to $\alpha$ and $\theta$. We pay special attention to the distribution of phase $\theta$ and reveal an interference-like pattern for the entanglement and purity quantifiers. In Section~\ref{section-kerr} we further apply the developed theory to a pump beam modulated by the Kerr nonlinear crystal. In Section~\ref{section-quadratures}, we study the effect of incoherent pump on the experimentally measurable variance of the combined quadratures. Section~\ref{section-conclusions} provides a summary and conclusions.

\section{Density operator for idler and signal modes} \label{section-density}

The interaction between the pump mode ($p$) and the down-converted modes in the non-degenerate OPG is described by the Hamiltonian (in units such that the Planck constant $\hbar=1$)
\begin{equation}\label{Hamiltonian}
 {H}_{\rm{int}} = g \left( {a}_{p} {a}^{\dagger}_{i}  {a}^{\dagger}_{s}  + {a}^{\dagger}_{p} {a}_{i}  {a}_{s} \right).
\end{equation}

\noindent where $g$ is the interaction constant related with the second-order electric susceptibility $\chi^{(2)}$, $\chi^{(2)} \ll \chi^{(1)}$. The interaction time $t$ between the pump mode and the down-converted modes is defined by the length of the nonlinear crystal. In realistic conditions, the product $gt \ll 1$ is a small parameter. This implies that the average number of photons in the pump beam, $\overline{N}$ is much greater than the average number of photons in the down-converted modes ($\sim g^2 t^2 \overline{N}$), so there is no depletion of the pump. 

\subsection{Coherent pump}

Suppose the pump is initially in the coherent state $\ket{\alpha}$, i.e. $a_p \ket{\alpha} = \alpha \ket{\alpha}$, and the idler and signal modes are in the vacuum state $\ket{0_i 0_s}$. The time evolution operator $U_t = \exp(-i H_{\rm int} t)$ with the Hamiltonian~\eqref{Hamiltonian} transforms the initial state into $\ket{\psi_{\alpha}(t)} = U_t \ket{\alpha}\ket{0_{i}0_{s}}$. The density operator of the idler and signal modes $\varrho_{is}(t)$ is obtained from $\ket{\psi_{\alpha}(t)}$ by taking the partial trace over the pump mode~\cite{Luchnikov2019}, $\varrho_{is}(t) = {\rm tr}_p {\varrho}\left(t\right)$. The density operator $\varrho_{is}(t)$ is always a linear combination of operators $\ket{m_i m_s} \bra{n_i n_s}$. The explicit form of $\ket{\psi_{\alpha}(t)}$ and $\varrho_{is}(t)$ is rather involved, so we consider two approaches: (i) perturbation theory with respect to the parameter $gt|\alpha| \lesssim 1$, (ii) parametric approximation that effectively replaces the operators $a_p$ and $a_p^{\dag}$ by c-numbers $\alpha$ and $\alpha^{\ast}$, respectively (see, e.g., the review~\cite{Hillery2009}).

\subsubsection{Perturbation theory}

As the experimentally achievable values $gt|\alpha| = 0.71 \div 0.97$~\cite{wang-2010,yan-2012,zhou-2015,yu-2016}, the main physical effects can be roughly illustrated by the perturbative approach, within which we derive formulas up to the second order of the parameter $gt$. We have
\begin{eqnarray}\label{psi-alpha}
&& \ket{\psi_{\alpha}\left(t\right)} = \left( I - it {H}_{\rm{int}}-\frac{t^{2}}{2} {H}_{\rm{int}}^2 + \ldots \right) \ket{\alpha}\ket{0_{i}0_{s}} \nonumber\\
&& = \ket{\alpha}\ket{0_{i}0_{s}}-igt\alpha\ket{\alpha}\ket{1_{i}1_{s}} \nonumber\\
&& - \frac{\alpha g^2t^2}{2}\left(2\alpha\ket{\alpha}\ket{2_{i}2_{s}}+\ket{\phi_{\alpha}}\ket{0_{i}0_{s}}\right) + o(g^2 t^2), 
\end{eqnarray}

\noindent where $\ket{\phi_{\alpha}} =  {a}_{p}^{\dagger} \ket{\alpha}$. The subscript $\alpha$ in $\ket{\psi_{\alpha}(t)}$ indicates that initial state of the pump is a pure coherent state $\ket{\alpha}$. The three-mode density operator of the pump, the idler, and the signal reads
\begin{eqnarray}
&& \!\!\!\!\!\!\!\!\!\! \ket{\psi_{\alpha}\left(t\right)}\bra{\psi_{\alpha}\left(t\right)} = \ket{\alpha}\bra{\alpha}
\nonumber\\
&& \!\!\!\!\!\!\!\!\!\! \otimes \big(
\ket{0_{i}0_{s}}\bra{0_{i}0_{s}} +  igt\alpha^{\ast}\ket{0_{i}0_{s}}\bra{1_{i}1_{s}}  -  g^2t^2(\alpha^{\ast})^{2}\ket{0_{i}0_{s}}\bra{2_{i}2_{s}} \nonumber\\
&& \!\!\!\!\!\!\!\!\!\! -igt\alpha\ket{1_{i}1_{s}}\bra{0_{i}0_{s}} + g^2t^2|\alpha|^2\ket{1_{i}1_{s}}\bra{1_{i}1_{s}} \nonumber\\
&& \!\!\!\!\!\!\!\!\!\! -g^2t^2\alpha^{2}\ket{2_{i}2_{s}}\bra{0_{i}0_{s}} \big) \nonumber\\
&& \!\!\!\!\!\!\!\!\!\! - \frac{g^2t^2}{2}\left(\alpha\ket{\phi_{\alpha}}\bra{\alpha}+\alpha^{\ast}\ket{\alpha}\bra{\phi_{\alpha}}\right) \otimes \ket{0_{i}0_{s}}\bra{0_{i}0_{s}} + o(g^2 t^2). \nonumber\\
\end{eqnarray}

The partial trace over the pump yields the following density matrix $\varrho_{is}^{\alpha}$ in the subspace of vectors $\{\ket{0_{i}0_{s}},\ket{1_{i}1_{s}},\ket{2_{i}2_{s}}\}$:
\begin{equation} \label{rho-is-alpha-perturbation}
\varrho_{is}^{\alpha} = \begin{pmatrix}
1-g^2t^2|\alpha|^2 & igt\alpha^{\ast} & -g^2t^2(\alpha^{\ast})^2 \\
-igt\alpha & g^2t^2|\alpha|^2 & o(g^2t^2)\\
-g^2t^2\alpha^2 & o(g^2t^2) & o(g^2t^2)
\end{pmatrix}.
\end{equation}

\subsubsection{Parametric approximation}

In the conventional parametric approximation with the pump state initially in a coherent state $\ket{\alpha}$, $\alpha = |\alpha|e^{i\theta}$, the initial state $\ket{\alpha}\ket{0_i 0_s}$ evolves into $\ket{\alpha}\ket{\psi_{is}}$, where
\begin{equation}\label{two-mode-squeezed}
\ket{\psi_{is}} = \sqrt{1-|\lambda (\alpha)|^2}\sum_{n=0}^{\infty}\lambda^{n}(\alpha)\ket{n_{i}n_{s}},
\end{equation}

\noindent is a two-mode squeezed vacuum state (see, e.g., \cite{Weedbrook}) with
\begin{equation} \label{lambda}
\lambda(\alpha) = -i e^{i\theta}\tanh{gt|\alpha|} = -i \frac{\alpha}{|\alpha|} \tanh{gt|\alpha|}.
\end{equation}

This approximation is valid for the intense pump with $|\alpha| \gg 1$, $gt \ll 1$, the average number of photons in the idler and signal modes $\bra{\psi_{is}} (a_i^{\dag}a_i + a_s^{\dag}a_s) \ket{\psi_{is}} = 2 \sinh^2(gt|\alpha|) \ll |\alpha|^2$, and $gt \exp(4 gt|\alpha|) \ll 1$~\cite{HILLERY,Hillery1984}. Clearly, the range of applicability of the parametric approximation is much wider than that of the perturbation theory.

The density matrix $\varrho_{is}$ reads
\begin{eqnarray} \label{rho-is-alpha-gpa}
\varrho_{is}^{\alpha} &=& (1 - |\lambda(\alpha)|^2) \sum_{n,m=0}^{\infty} \lambda^{n}(\alpha) [\lambda^{m}(\alpha)]^{\ast} \ket{n_{i}n_{s}}\bra{m_{i}m_{s}} \nonumber\\
&=& \sum_{n,m=0}^{\infty} \frac{(-ie^{i\theta})^{n-m} \tanh^{n+m}gt|\alpha|}{\cosh^2gt|\alpha|} \ket{n_{i}n_{s}}\bra{m_{i}m_{s}}.\nonumber\\
\end{eqnarray}

\subsection{Incoherent pump}

For a general initial pump state $\varrho_p$ we use the diagonal sum representation $\varrho_p = \int P(\alpha) \ket{\alpha}\bra{\alpha} d^2 \alpha$, where $P(\alpha)$ is the Glauber-Sudarshan function~\cite{GLAUBER,Sudarshan}. Exploiting the linearity of the quantum evolution, we get the three-mode density operator of the pump, idler, and signal at time $t$
\begin{eqnarray}\label{do3-general}
{\varrho}(t) &=& U_t \ \varrho_p \otimes \ket{0_{i}0_{s}}\bra{0_{i}0_{s}} \ U_t^{\dag} \nonumber\\
 &=& \int P\left(\alpha \right) \ket{\psi_{\alpha}\left(t\right)}\bra{\psi_{\alpha}\left(t\right)} d^2 \alpha .
\end{eqnarray}

Taking partial trace over the pump, we get
\begin{equation} \label{rho-is-general-P}
\varrho_{is} = \int P\left(\alpha \right) \varrho_{is}^{\alpha} d^2 \alpha .
\end{equation}

\subsubsection{Perturbation theory}

Substituting~\eqref{rho-is-alpha-perturbation} for $\varrho_{is}^{\alpha}$ in Eq.~\eqref{rho-is-general-P}, we get
the following matrix representation of $\varrho_{is}$ in the subspace of vectors $\{\ket{0_{i}0_{s}},\ket{1_{i}1_{s}},\ket{2_{i}2_{s}}\}$:
\begin{equation} 
\label{FIN_DM}
\varrho_{is} = \begin{pmatrix}
1-g^2t^2c_{11} & igtc_{01} & -g^2t^2c_{02}\\
-igtc^{\ast}_{01} & g^2t^2c_{11} & o(g^2t^2)\\
-g^2t^2c^{\ast}_{02} & o(g^2t^2) & o(g^2t^2)
\end{pmatrix},
\end{equation}

\noindent where the coefficients $c_{mn}$ read
\begin{eqnarray}\label{c-coefficients}
c_{mn} &=& \int P(\alpha) \alpha^m (\alpha^{\ast})^n d^2 \alpha  \nonumber\\
&=& \int\limits_0^{\infty} d|\alpha| \int\limits_{0}^{2\pi} d\theta P\left(|\alpha|e^{i\theta}\right) |\alpha|^{m+n+1} e^{i\theta(m-n)}. \quad
\end{eqnarray}

\noindent Note that $c_{00} = 1$ because $\int P(\alpha) d^2 \alpha = {\rm tr} [\varrho_p] = 1$. 

\subsubsection{Generalized parametric approximation}

Ref.~\cite{Hillery1995} generalizes the parametric approximation to the case of pure initial pump states $\ket{\psi_p} = \frac{1}{\pi} \int \ip{\alpha}{\psi_p} \ket{\alpha} d^2\alpha$. Each coherent constituent $\ket{\alpha}$ in the pump results in the signal-idler field given by Eq.~\eqref{two-mode-squeezed}, so the integration of Eq.~\eqref{two-mode-squeezed} with the kernel $\frac{1}{\pi}\ip{\alpha}{\psi_p}$ provides the signal-idler output state for any pure pump $\ket{\psi_p}$. 

We follow the same idea for a general (mixed) pump density operator $\varrho_p = \int P(\alpha) \ket{\alpha}\bra{\alpha} d^2\alpha$, where $P(\alpha)$ is the Glauber-Sudarshan function of the pump field. Substituting \eqref{rho-is-alpha-gpa} for $\varrho_{is}^{\alpha}$ in Eq.~\eqref{rho-is-general-P}, we get the resulting density operator $\varrho_{is}$ of the idler and signal modes. $\varrho_{is}$ is a mixture of states \eqref{rho-is-alpha-gpa} with weights $P(\alpha)$ that can be negative in general, namely,
\begin{eqnarray}
&& \varrho_{is} = \sum_{n,m=0}^{\infty} \varrho_{is}^{nm} \ket{n_{i}n_{s}}\bra{m_{i}m_{s}}, \label{dmatrix-two-mode} \\
&& \varrho_{is}^{nm} = \int P(\alpha)(1 - |\lambda(\alpha)|^2) \lambda^{n}(\alpha) [\lambda^{m}(\alpha)]^{\ast} d^{2}\alpha \nonumber\\
&& = \int\limits_0^{\infty} |\alpha| \, d|\alpha| \int\limits_0^{2\pi} d\theta P(|\alpha|e^{i\theta}) \frac{ (-ie^{i\theta})^{n-m} \tanh^{n+m}gt|\alpha|}{\cosh^2gt|\alpha|}. \nonumber\\ \label{rho-n-m}
\end{eqnarray}

Eq.~\eqref{dmatrix-two-mode} provides a solution of the problem in the generalized parametric approximation with no truncation in the Fock space. This is in contrast to the perturbative solution~\eqref{FIN_DM} that involves the truncation of the Fock space up to 2 photons in the signal-idler field.

The validity of the generalized parametric approximation is
considered by using a path-integral representation of the coherent-state propagator in Ref.~\cite{Hillery1984}. The conditions under which this approximation is justified are specified in Ref.~\cite{Hillery1984} for the degenerate parametric amplifier with the signal mode initially in the vacuum state. Ref.~\cite{Hillery1995} extends the ideas of Ref.~\cite{Hillery1984} to the case of the non-degenerate parametric amplifier. Following~\cite{Hillery1984,Hillery1995}, we expect the generalized parametric approximation to be valid if the average number of photons in the pump beam ${\rm tr}[\varrho_p a_p^{\dag} a_p] \gg 1$ (high pump intensity), ${\rm tr}[\varrho_{is} (a_i^{\dag}a_i + a_s^{\dag}a_s)] \ll {\rm tr}[\varrho_p a_p^{\dag} a_p]$ (no pump depletion), $gt \ll 1$, and $gt \exp\left(4 gt \sqrt{{\rm tr}[\varrho_p a_p^{\dag} a_p]} \right) \ll 1$ (the propagator is a slowly varying function of the outcome pump amplitude). In terms of the $P$-function these conditions read
\begin{eqnarray}
&& \int P(\alpha) |\alpha|^2 d^2 \alpha \gg 1, \label{condition-GPA-first} \\
&& gt \ll 1,\\
&& 2 \int P(\alpha) \sinh^2(gt|\alpha|) d^2 \alpha \ll \int P(\alpha) |\alpha|^2 d^2 \alpha, \\
&& gt \exp\left(4 gt \sqrt{\int P(\alpha) |\alpha|^2 d^2 \alpha } \right) \ll 1. \label{condition-GPA-final} 
\end{eqnarray}

Note that the coefficients~\eqref{rho-n-m} contain all orders of the small parameter $gt$ in contrast to the perturbation theory restricted by the second order of $gt$. In fact, conditions~\eqref{condition-GPA-first}--\eqref{condition-GPA-final}  are much less restrictive than the condition $gt\sqrt{{\rm tr}[\varrho_p a_p^{\dag} a_p]} \lesssim 1$ for the validity of the perturbative approach. Therefore, the generalized parametric approximation represents a definite improvement over the perturbative result and enables us to study the effect of incoherent pump at a longer timescale.

\section{Purity and entanglement of the idler-signal field} \label{section-purity}

The purity parameter ${\rm tr}[\varrho_{is}^2]$ quantifies how close the given state $\varrho_{is}$ is to a pure one. Purity of a continuous-variable quantum state can be operationally calculated via homodyne measurements~\cite{bellini-2012}. Note that ${\rm tr}[\varrho_{is}^2] = 1$ if and only if $\varrho_{is}$ is pure. The related quantifier is the linear entropy 
\begin{equation}\label{PURITY}
S_{L} = 1 - {\rm tr}[\varrho_{is}^{2}].
\end{equation}

\noindent The greater the linear entropy, the more mixed the state is. 

Within the perturbation theory, the linear entropy of the state~\eqref{FIN_DM} equals
\begin{equation}\label{S_L}
S_L = 2g^2t^2 \left(c_{11}-|c_{01}|^2\right) + o(g^2t^2)
\end{equation}

\noindent and grows quadratically with time $t$ while $g^2 t^2 {\rm tr}[\varrho_p a_p^{\dag} a_p] \ll 1$ (short timescale). 

Within the generalized parametric approximation we get the linear entropy for a longer timescale,
\begin{equation} \label{S-gpa}
S_L = 1 - \sum_{n,m=0}^{\infty} |\varrho_{is}^{nm}|^2,
\end{equation}

\noindent which strongly depends on $P(\alpha)$.

Since $\varrho_{is}$ is a linear combination of operators $\ket{n_i n_s} \bra{m_i m_s}$, its entanglement can be effectively quantified via the negativity measure~\cite{Vidal}
\begin{equation}\label{NEGATIVITY}
N = \frac{ \| \varrho_{is}^{T_{s}} \|_{1} - 1}{2},
\end{equation}

\noindent where $\|A\|_1 = \sum_k |\lambda_k|$ for a Hermitian operator $A$ with spectrum $\{\lambda_k\}$, $T_s$ is a partial transposition in the signal mode, i.e., $\left( \ket{n_i n_s} \bra{m_i m_s} \right)^{T_s} = \ket{n_i m_s} \bra{m_i n_s}$. The physical meaning of partial transposition is twofold: It can be seen as a mirror reflection in the phase space of the corresponding mode~\cite{simon-2000} and as a local time reversal fot that mode~\cite{sanpera-1997}. Negativity $N$ has an operational meaning too: the entanglement cost for the exact preparation of the quantum state $\varrho$ using quantum operations preserving the positivity of the partial transpose is bounded from below by $\log_2(2N+1)$~\cite{audenaert-2005}.

Provided the two-mode state is separable, the partial transposition would not affect the positivity of the density matrix, otherwise it may transform the density operator into a non-positive operator, which is an indication of entanglement. Thus, the negativity vanishes for separable states $\varrho_{is} = \sum_{k} p_k \varrho_i^{(k)} \otimes \varrho_s^{(k)}$, whereas $N > 0$ indicates the state $\varrho_{is}$ is entangled. In the following equations, nonzero elements of $\varrho_{is}$ in the basis $\ket{m_i n_s}$ are marked by black squares in Eq.~\eqref{PPT-matrix} below. The transposition with respect to sybsystem $s$ moves the elements of $\varrho_{is}$ to positions marked by empty squares in Eq.~\eqref{PPT-matrix}. Dotted lines in Eq.~\eqref{PPT-matrix} denote $2 \times 2$ minors $\left( \begin{array}{cc} 0 & \square\\ \square & 0 \end{array} \right)$ with negative eigenvalue $- |\square|$.

\begin{equation} \label{PPT-matrix} \setstretch{2.0} {\scriptsize
 \bordermatrix{ & 00 & 01 & 02 & \ldots & \VR & 10 & 11 & 12 & \ldots & \VR & 20 & 21 & 22 & \ldots \cr
    00 &                 \blacksquare &  &  &  & \VR &  & \rn{0011}{\blacksquare} &  &  & \VR &  &  & \rn{0022}{\blacksquare} &  \cr
    01 &                  & \tikzmark{left1} &  &  & \VR & \rn{0110}{\square} &  &  &  & \VR &  &  &  &  \cr
    02 &                  &  & \tikzmark{left2} &  & \VR &  &  &  &  & \VR & \rn{0220}{\square} &  &  &  \cr
    \ldots &              &  &  &  & \VR &  &  &  &  & \VR &  &  &  &  \cr
  \noalign{\smallskip\hrule\smallskip}
    10 &                  & \rn{1001}{\square} &  &  & \VR &  \tikzmark{right1}  &  &  &  & \VR &  &  &  & \cr
    11 &                 \rn{1100}{\blacksquare} &  &  &  & \VR &  & \blacksquare &  &  & \VR &  &  & \rn{1122}{\blacksquare} &  \cr
    12 &                  &  &  &  & \VR &  &  & \tikzmark{left3} &  & \VR &  & \rn{1221}{\square} &  &  \cr
    \ldots &             &  &  &  & \VR &  &  &  &  & \VR &  &  &  &  \cr
  \noalign{\smallskip\hrule\smallskip}
    20 &                 &  & \rn{2002}{\square} &  & \VR &  &  &  &  & \VR & \tikzmark{right2} &  &  &  \cr
    21 &                 &  &  &  & \VR &  &  & \rn{2112}{\square} &  & \VR &  & \tikzmark{right3} &  &  \cr
    22 &                \rn{2200}{\blacksquare} &  &  &  & \VR &  & \rn{2211}{\blacksquare} &  &  & \VR &  &  & \blacksquare &  \cr
    \ldots &             &  &  &  & \VR &  &  &  &  & \VR &  &  &  &  \cr
}}
\end{equation}
\DrawBox[thick, red, dotted]{left1}{right1}
\DrawBox[thick, red, dotted]{left2}{right2}
\DrawBox[thick, red, dotted]{left3}{right3}

\begin{tikzpicture}[overlay,remember picture]
\draw [->] (0011) -- (0110);
\draw [->] (1100) -- (1001);
\draw [->] (0022) -- (0220);
\draw [->] (2200) -- (2002);
\draw [->] (1122) -- (1221);
\draw [->] (2211) -- (2112);
\end{tikzpicture}

Analytically, $\varrho_{is}^{T_{s}} = \sum_{n,m=0}^{\infty} \varrho_{is}^{nm} \ket{n_{i} m_{s}}\bra{m_{i} n_{s}}$. Choose $n' \neq m'$ and consider a $2 \times 2$ submatrix of $\varrho_{is}^{T_{s}}$ obtained by deleting all rows and columns except those with multiindices $n'm'$ and $m'n'$. Examples of such submatricies are shown in Eq.~\eqref{PPT-matrix} by dotted lines. As $n' \neq m'$, the diagonal elements of such submatrices vanish, i.e., $\bra{n_{i}' n_{s}'} \varrho_{is}^{T_{s}} \ket{n_{i}' n_{s}'} = \bra{m_{i}' m_{s}'} \varrho_{is}^{T_{s}} \ket{m_{i}' m_{s}'} = 0$. Determinant of such a submatrix equals $-|\varrho_{is}^{n'm'}|^2 \leq 0$ and represents a principal minor of $\varrho_{is}^{T_{s}}$. By Sylvester's criterion, a Hermitian matrix is positive-semidefinite if and only if all its principal minors are nonnegative. Therefore, $\varrho_{is}^{T_{s}}$ is not positive-semidefinite unless $\varrho_{is}^{n'm'} = 0$ for all $n' \neq m'$. Hence, by the Peres--Horodecki criterion~\cite{peres-1996,horodecki-1996} $\varrho_{is}$ is entangled if $\sum_{m' < n'} |\varrho_{is}^{n'm'}| > 0$. On the other hand, if $\sum_{m' < n'} |\varrho_{is}^{n'm'}| = 0$, then $\varrho_{is}$ is diagonal and separable. Finally, $\varrho_{is}$ is entangled if and only if $\sum_{m' < n'} |\varrho_{is}^{n'm'}| > 0$. The latter quantity is nothing else but the negativity. In fact, negative eigenvalues of the partially transformed state $\varrho_{is}^{T_{s}}$ are negative eigenvalues of $2 \times 2$ submatrices described above. Summing them, we get the negativity
\begin{equation} \label{N-general}
N = \sum_{m < n} |\varrho_{is}^{nm}|.
\end{equation}

\noindent The negativity of the studied state $\varrho_{is}$ vanishes if and only if all non-diagonal elements of $\varrho_{is}$ are equal to zero. In other words, the feature of $\varrho_{is}$ is that it is entangled if and only if $N > 0$. Formula~\eqref{N-general} enables one to calculate the negativity by learning the elements of the density matrix, e.g., via the homodyne tomography~\cite{raymer-1996,kurochkin-2014}. 

In the perturbation theory, for the state~\eqref{FIN_DM} we get 
\begin{equation}\label{N_P}
N = gt|c_{01}|+g^2t^2|c_{02}|+o(g^2t^2).
\end{equation}

\noindent The negativity $N$ grows linearly with time $t$ while $g^2 t^2 {\rm tr}[\varrho_p a_p^{\dag} a_p] \ll 1$ (short timescale).
For a longer timescale one should use Eq.~\eqref{rho-n-m} for coefficients $\varrho_{is}^{nm}$, which in turn are expressed through the Glauber-Sudarshan function $P(\alpha)$ of the pump state. 

At the end of this section, we discuss the entanglement and purity of the state $\varrho_{is}$ produced with the help of the pure coherent pump. In the following sections, we discuss particular models of the incoherent pump and its effect on the idler-signal entanglement.

\subsection{Coherent pump}

Let $\varrho_p = \ket{\alpha_0}\bra{\alpha_0}$, then $P(\alpha) = \delta(\alpha - \alpha_0)$. In the perturbation theory up to the second order of $gt$, we get
\begin{equation}\label{le-neg-coherent}
S_{L} = 0 + o(g^2t^2), \quad N = gt|\alpha_{0}|+(gt|\alpha_{0}|)^2 + o(g^2t^2).
\end{equation}

In the parametric approximation, the output state of the signal and idler is the pure state~$\psi_{is}$ with $\alpha = \alpha_0$ ($S_L = 0$). In the generalized parametric approximation, we get the following expression for the negativity:
\begin{eqnarray}
N &=& (1-|\lambda(\alpha_0)|^2) \sum_{n<m}^{\infty}
|\lambda(\alpha_0)|^{n+m} =
\frac{|\lambda(\alpha_0)|}{1-|\lambda(\alpha_0)|} \nonumber\\
&=& \frac{\tanh gt|\alpha_0|}{1 - \tanh gt|\alpha_0|} = \frac{1}{2}\left( e^{2gt|\alpha_0|} - 1 \right) \label{negativity-coherent-gpa} \\
&=& gt|\alpha_0| + g^2 t^2 |\alpha_0|^2 + \frac{2}{3} g^3 t^3 |\alpha_0|^3 + o(g^3 t^3 |\alpha_0|^3). \nonumber
\end{eqnarray}

\noindent In contrast, extending the exact formula~\eqref{psi-alpha} to the third order with respect to $gt$ and calculating the negativity of the exact $\varrho_{is}$, we get
\begin{equation}
N = gt|\alpha_0| + g^2 t^2 |\alpha_0|^2 + g^3 t^3 \left( \frac{2}{3} |\alpha_0|^3 - \frac{|\alpha_0|}{6} \right) + o(g^3 t^3|\alpha_0|^3),
\end{equation}

\noindent which negligibly differs from Eq.~\eqref{negativity-coherent-gpa} if $g^3 t^3|\alpha_0| \ll 1$. The latter inequality is an immediate implication of the additional inequality $gt \exp(4 gt|\alpha_0|) \ll 1$ and the inequality $gt \ll 1$, which are necessary for justification of the parametric approximation~\cite{Hillery1984}. 

Based on this example, we conclude that the quantity $g^2 t^2 \left\{ \exp \left( 4 g t \sqrt{{\rm tr}[\varrho_p a_p^{\dag} a_p]} \right) - 1 \right\}$ is an estimate for the trace distance between the exact solution for $\varrho_{is}$ and the approximate solution~\eqref{dmatrix-two-mode}--\eqref{rho-n-m}.

\section{Thermal pump} \label{section-thermal}

A rather intense radiation for the pump is produced not only by a laser but also by alternative optical sources such as amplified spontaneous emission (ASE) and light--emitting diodes (LED). In contrast to the laser radiation, the ASE and LED produce a light of low spatial and temporal coherence~\cite{tziraki-2000,di-lorenzo-pires-2010,boitier-2009,jechow-2013}. In particular, Ref.~\cite{boitier-2009} reports a doped fibre ASE radiation with the degree of second-order coherence $g^{(2)}(0)=1.97 \pm 0.1$, which is higher than that for the halogen lamp ($g^{(2)}(0)=1.8 \pm 0.1$) exhibiting blackbody characteristics at 3000~K. Similarly, the superluminescent diode emits a completely incoherent light which has the same characteristics as the thermal radiation~\cite{lee-1973,jechow-2013,kurzke-2017,hartmann-2017}. A pseudothermal light of high intensity is also generated by inserting a spinning glass diffuser (Arecchi’s wheel) in the path of the laser light~\cite{Arecchi,Valencia,Parigi,Zavatta,vidrighin-2016}. We describe the completely incoherent (thermal) radiation in the selected pump mode by the density operator $\varrho_p = \int P(\alpha) \ket{\alpha}\bra{\alpha} d^2 \alpha$ with the Glauber-Sudarshan function
\begin{equation} \label{thermal}
P\left(\alpha \right) = \frac{1}{\pi\overline{n}} \exp \left( -\frac{|\alpha|^2}{\overline{n}} \right), \end{equation}

\noindent where $\overline{n}$ is the average number of photons in the pump mode. 

As $P(\alpha)$ does not depend on $\theta = {\rm arg}\alpha$, the expressions~\eqref{c-coefficients} and~\eqref{rho-n-m} vanish if $n \neq m$. Consequently, in both the perturbation theory and the generalized parametric approximation, the negativity of $\varrho_{is}$ vanishes too, i.e,
\begin{equation}\label{Niszero}
N=0.
\end{equation}

\noindent This means that there is no entanglement between the idler mode and the signal mode. Instead, only classical correlations are present as $\varrho_{is} = \sum_{n=0}^{\infty} p_n \ket{n_i n_s} \bra{n_i n_s}$, where $p_n$ is the probability to observe $n$ photons in either of idler or signal modes. Although the photon pairs are still created simultaneously in such an OPG, this form of correlations can be created via local operations and classical communications~\cite{Horodecki-2009}.

To evaluate the purity of $\varrho_{is}$, we calculate the coefficients 
\begin{equation}
c_{01} = 0, \qquad c_{11}=\overline{n}, \qquad c_{02} = 0
\end{equation}

\noindent in the perturbation theory, which results in 
\begin{equation}\label{le-neg-thermal}
S_{L} = 2\overline{n}\left(gt\right)^2 + o(g^2t^2). 
\end{equation}

The generalized parametric approximation is valid if $\overline{n}\gg1, gt\ll1, gt\exp{\left(4gt\sqrt{\overline{n}}\right)}\ll1$. We find the lower bound on coefficients $\varrho_{is}^{mm}$ by using the inequalities $1/\cosh^2 x \geq e^{-x^2}$ and $\tanh x \geq x e^{-x^2}$
in Eq.~\eqref{rho-n-m}. The result is
\begin{eqnarray}
&& \varrho_{is}^{mm} \geq \frac{m!}{g^2 t^2 \overline{n} \left( 2m+1 + \frac{1}{g^2 t^2 \overline{n}}\right)^{m+1}}, \\
&& S_L \leq 1 - \sum_{m=0}^{\infty} \frac{(m!)^2}{g^4 t^4 \overline{n}^2 \left( 2m+1 + \frac{1}{g^2 t^2 \overline{n}}\right)^{2m+2}}. \label{S-L-thermal}
\end{eqnarray}

Let us summarize the results of this section. The thermal pump of arbitrarily high intensity is not able to produce entanglement between the signal and the idler. This is due to a fact that the phase distribution for such a radiation is uniform. Only classical correlations are present in the signal-idler field in this case. The signal-idler field is mixed and its linear entropy is bounded from below by Eq.~\eqref{le-neg-thermal} and from above by Eq.~\eqref{S-L-thermal}.

\section{Noisy coherent pump} \label{section-noisy-coherent}

\subsection{Phase-insensitive Gaussian noise}

The situation is different for such pump states that still exhibit coherent properties. For instance, the superluminescent diode with a balance between spontaneous and stimulated emission~\cite{blazek-2011} and the superluminescent diode with controlled optical feedback~\cite{kiethe-2017} produce the partially coherent light with $1 < g^{(2)}(0) < 2$. Such a radiation represents a superposition of the thermal radiation and the coherent radiation in the sense of Ref.~\cite{GLAUBER}, formula 7.19. In other words, such a radiation describes the action of the phase-insensitive  Gaussian noise on the pure coherent state and is known as a displaced thermal state or a noisy coherent state~\cite{zhao-2017}. This state is also experimentally simulated by superimposing the thermal light and the coherent light on a beamsplitter~\cite{bondani-2009}. The $P$-function of a displaced thermal state reads
\begin{equation} \label{displaced-thermal}
P\left( \alpha \right) = \frac{1}{\pi \overline{n}} \exp \left( - \frac{|\alpha - \alpha_{0}|^2}{\overline{n}}\right).
\end{equation}

\begin{figure}
\includegraphics[width=8cm]{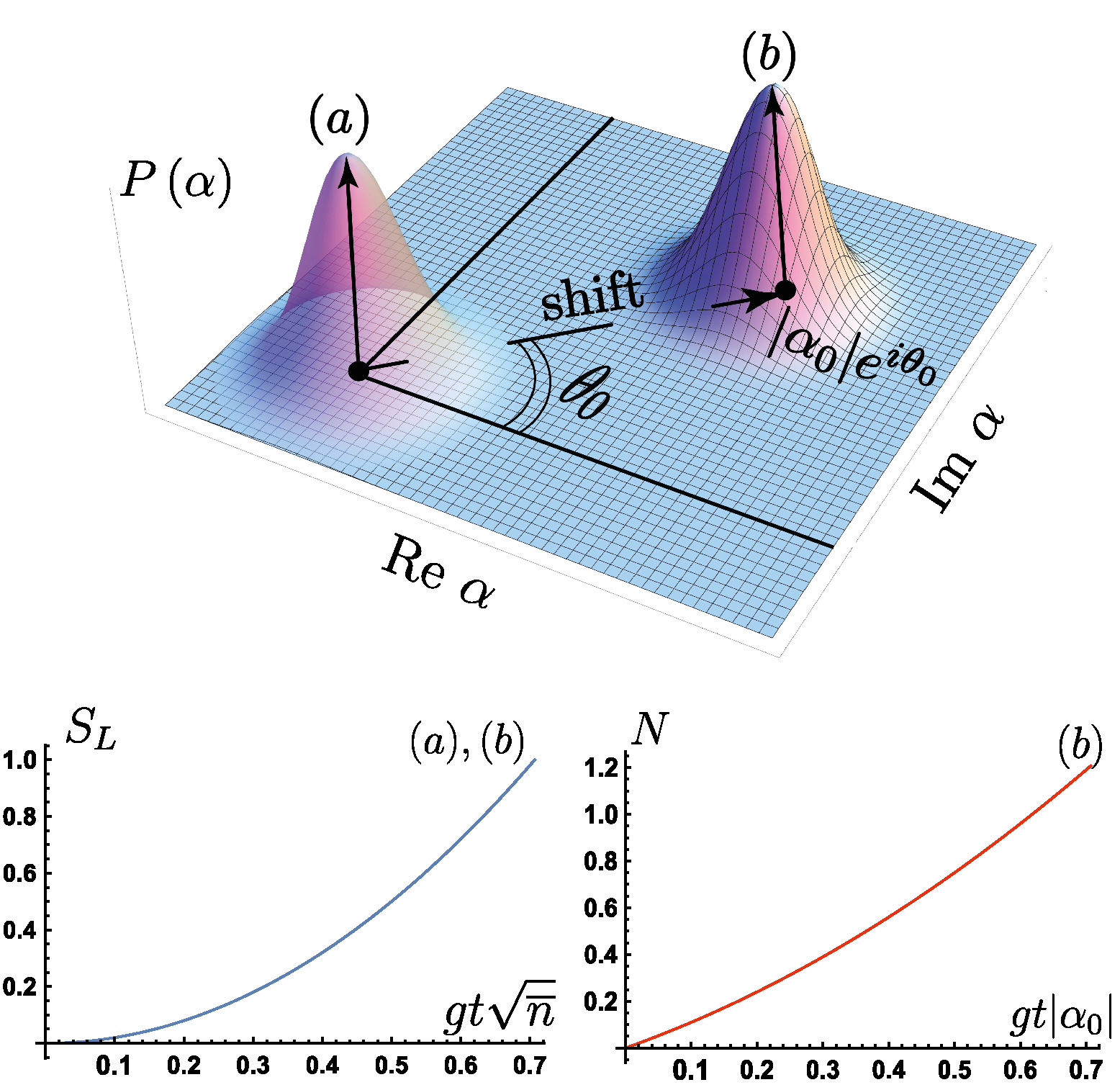}
\caption{Properties of the thermal pump (a) and the displaced thermal pump (b): the $P$-function in the phase space (top); the linear entropy of the idler and signal modes (bottom left); the negativity measure of entanglement for the idler and signal modes (bottom right). Negativity vanishes for the thermal pump, Eq.~\eqref{Niszero}. }
\label{figure-2}
\end{figure}

\noindent Geometrically, the state~\eqref{displaced-thermal} is obtained from the thermal state~\eqref{thermal} by a shift in the phase space $({\rm Re}\alpha, {\rm Im}\alpha)$; the displacement equals $|\alpha_0|$ and its direction is determined by the angle $\theta_0 = {\rm arg} \alpha_0$, see Fig.~\ref{figure-2}.

The use of the displaced thermal light~\eqref{displaced-thermal} as a pump results in the state of idler and signal modes~\eqref{FIN_DM} with the following parameters in the perturbation theory:
\begin{equation}
c_{01} = |\alpha_{0}|e^{-i\theta_{0}}, \qquad c_{11}=|\alpha_{0}|^2+\overline{n}, \qquad c_{02} = |\alpha_{0}|^2e^{-i2\theta_{0}}.
\end{equation}

The linear entropy and negativity of the idler-signal state in the perturbation theory read
\begin{equation}\label{le-neg-displaced-thermal}
S_{L} = 2\overline{n}\left(gt\right)^2 + o(g^2t^2), \quad N = gt|\alpha_{0}|+(gt|\alpha_{0}|)^2 + o(g^2t^2).
\end{equation}

The idler and signal modes are entangled and the degree of entanglement is proportional to the displacement $|\alpha_0|$ if $gt|\alpha_0| \ll 1$. The greater $\overline{n}$, the more mixed the state $\varrho_{is}$ is. Therefore, the thermal contribution in the noisy coherent pump leads to a decrease of the purity of $\varrho_{is}$, whereas the coherent contribution in the noisy coherent pump determines the entanglement of $\varrho_{is}$. 

A comparison the idler-signal properties for the thermal pump and the displaced thermal pump is presented in Fig.~\ref{figure-2} for a short timescale ($gt|\alpha_0| \lesssim 1$).
The linear entropy is the same for both cases, whereas the negativity is different. These results are applicable to the experimental study of the OPG with the incoherent pump produced by a light--emitting diode~\cite{LIT1,LIT2,LIT3}.

To study the behavior at a longer timescale ($gt|\alpha_0| > 1$) we use the generalized parametric approximation. Substituting~\eqref{displaced-thermal} and~\eqref{lambda} into \eqref{dmatrix-two-mode}, we get
\begin{eqnarray}
\label{dmatrix-displaced-gen} && \varrho_{is}^{nm} = \frac{(-i)^{n-m}}{\pi \overline{n}} \int \frac{\alpha^n (\alpha^m)^{\ast} \tanh^{n+m} gt|\alpha|}{|\alpha|^{n+m} \cosh^2 gt|\alpha|} \nonumber\\
&& \qquad\qquad\qquad \times\exp\left( - \frac{|\alpha - \alpha_0|^2}{\overline{n}} \right) d^{2}\alpha.
\end{eqnarray}

\noindent The obtained expression is simplified if $\overline{n} \ll |\alpha_0|^2$, when one can use a two-dimensional analogue of the formula $\int f(x) e^{(x-y)^2/ \overline{n}}dx = \sqrt{\pi \overline{n}} f(y) + \frac{1}{4} \sqrt{\pi \overline{n}^3} f''(y) + o(\overline{n}^{3/2})$. We get
\begin{eqnarray}
&& \!\!\!\!\!\!\!\!\!\! \varrho_{is}^{nm} =  \frac{(-i)^{n-m} \alpha_0^n (\alpha_0^m)^{\ast} \tanh^{n+m} gt|\alpha_0|}{|\alpha_0|^{n+m} \cosh^2 gt|\alpha_0|}  \nonumber\\
&& \!\!\!\!\!\!\!\!\!\!  + \frac{\overline{n}}{4}(-i)^{n-m}  \left( \frac{\partial^2}{\partial ({\rm Re}\alpha)^2} + \frac{\partial^2}{\partial ({\rm Im}\alpha)^2} \right)  \nonumber\\
&& \!\!\!\!\!\!\!\!\!\! \times \left. \frac{\alpha^n (\alpha^m)^{\ast} \tanh^{n+m} gt|\alpha|}{|\alpha|^{n+m} \cosh^2 gt|\alpha|} \right\vert_{\alpha = \alpha_0} + o\left(\frac{\overline{n}}{|\alpha_0|^2}\right). \label{rho-n-m-gpa-displaced-thermal}
\end{eqnarray}

\noindent The first term in Eq.~\eqref{rho-n-m-gpa-displaced-thermal} corresponds to a pure coherent pump, and the second term is a deviation caused by the pump incoherence. If additionally $gt|\alpha_0| \lesssim 1$, then we calculate the negativity and the linear entropy of $\varrho_{is}$ up to the third order of $gt|\alpha_0|$. The linear entropy coincides with that in Eq.~\eqref{le-neg-displaced-thermal}, whereas the negativity reads:
\begin{eqnarray}
&& \label{N-gad-noisy-coherent} N = gt|\alpha_0| + g^2 t^2 |\alpha_0|^2 + \frac{2}{3} g^3 t^3 |\alpha_0|^3 \left( 1 - \frac{\overline{n}}{|\alpha_0|^2} \right)  \nonumber\\
&& \qquad + o(g^3 t^3 |\alpha_0|^3). \label{negativity-gpa-displaced-thermal}
\end{eqnarray}

\noindent The negativity~\eqref{negativity-gpa-displaced-thermal} is less than that in Eq.~\eqref{negativity-coherent-gpa}, which means the thermal noise in the pump decreases the degree of entanglement.

Let us summarize the results of this section. The admixture of the phase-insensitive Gaussian noise to a coherent pump results in the decrease of entanglement between the signal and the idler. This effect is revealed in the generalized parametric approximation, Eq.~\eqref{N-gad-noisy-coherent}, whereas it is concealed in the perturbative approach up to the second order of $gt$, Eq.~\eqref{le-neg-displaced-thermal}. The greater is the noise intensity, the less is the purity of the signal-idler field. 

\subsection{Phase-sensitive Gaussian noise}

Phase sensitive noise is typical for phase-sensitive linear amplifiers~\cite{Caves1982}. The $P$-function for a coherent pump subjected to the phase-sensitive Gaussian noise reads
\begin{eqnarray} \label{phase-sensitive-noisy-coherent}
P\left( \alpha \right) &=& \frac{1}{\pi \sqrt{\overline{n}_1 \overline{n}_2}} \exp \Bigg( - \frac{\{{\rm Re}[(\alpha - \alpha_{0})e^{-i\varphi}]\}^2}{\overline{n}_1} \nonumber\\
&& - \frac{\{{\rm Im}[(\alpha - \alpha_{0})e^{-i\varphi}]\}^2}{\overline{n}_2} \Bigg).
\end{eqnarray}

\begin{figure}
\includegraphics[width=8cm]{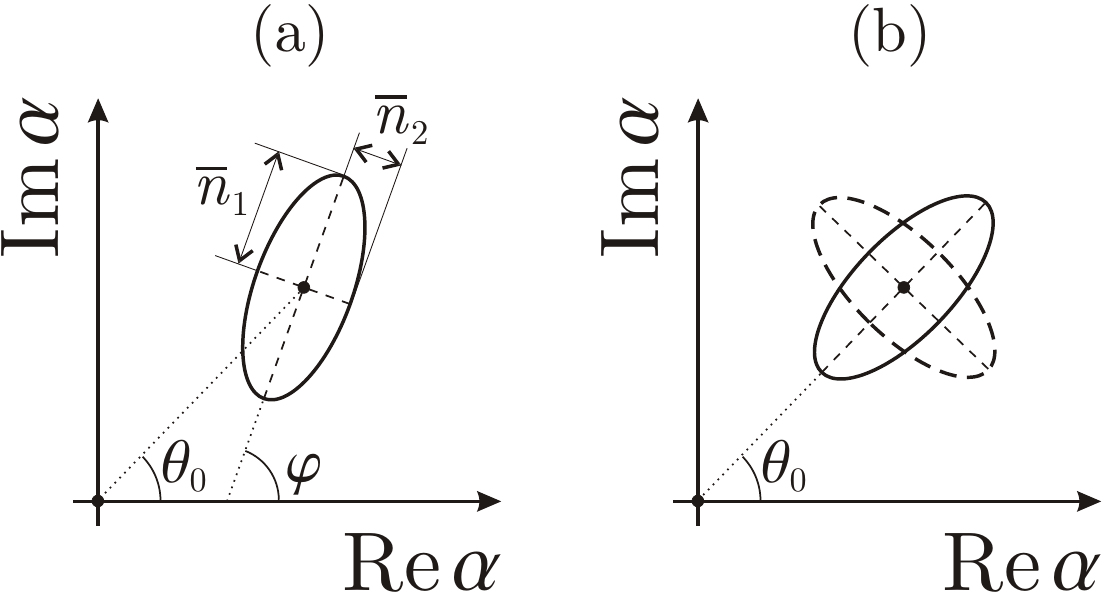}
\caption{(a) Phase-space schematic for admixture of a phase-sensitive Gaussian noise with parameters $\overline{n}_1$, $\overline{n}_2$, and $\varphi$ to a coherent pump $\ket{\alpha_0}$, $\alpha_0 = |\alpha_0|e^{i\theta_0}$. (b) Configuration of parameter $\varphi$ for the maximal increment in the signal-idler entanglement (solid line ellipse) and the maximal decrement in the signal-idler entanglement (dashed line ellipse).}
\label{figure-3}
\end{figure}

\noindent Geometrically, the state~\eqref{phase-sensitive-noisy-coherent} is obtained from the thermal state~\eqref{thermal} by squeezing the horizontal and vertical axes in the phase space $({\rm Re}\alpha, {\rm Im}\alpha)$ by factors $\overline{n}_1/\overline{n}$ and $\overline{n}_2/\overline{n}$, respectively, then rotating by the angle $\varphi$ around the origin of the phase space, and finally shifting by a vector $({\rm Re}\alpha_0, {\rm Im}\alpha_0)$, see Fig.~\ref{figure-3}(a).

The noisy pump~\eqref{phase-sensitive-noisy-coherent} results in the state of idler and signal modes~\eqref{FIN_DM} with the following parameters in the perturbation theory:
\begin{eqnarray}
&& c_{01} = |\alpha_{0}|e^{-i\theta_{0}}, \qquad c_{11}=|\alpha_{0}|^2 + \frac{\overline{n}_1 + \overline{n}_2}{2} \, , \\
&& c_{02} = |\alpha_{0}|^2e^{-i2\theta_{0}} + \frac{\overline{n}_1 - \overline{n}_2}{2} e^{- i 2 \varphi}.
\end{eqnarray}

The linear entropy and negativity of the idler-signal state in the perturbation theory read
\begin{eqnarray}
&& S_{L} = (\overline{n}_1 + \overline{n}_2) (gt)^2 + o(g^2t^2), \\
&& N = gt|\alpha_{0}| \nonumber\\
&& +(gt|\alpha_{0}|)^2 \sqrt{1 + \frac{\overline{n}_1 - \overline{n}_2}{|\alpha_0|^2} \cos 2(\theta_0 - \varphi) + \left( \frac{\overline{n}_1 - \overline{n}_2}{2 |\alpha_0|^2} \right)^2 } \nonumber\\
&& + o(g^2t^2). \label{N-phase-insensitive-noise}
\end{eqnarray}

If the noise $\overline{n}_1,\overline{n}_2 \ll |\alpha_0|^2$, then  $N \approx gt|\alpha_{0}| +(gt|\alpha_{0}|)^2 + g^2 t^2 \frac{\overline{n}_1 - \overline{n}_2}{2} \cos 2(\theta_0 - \varphi)$. Comparing this result with Eq.~\eqref{le-neg-displaced-thermal}, we conclude that the phase sensitive noise affects the negativity stronger than the phase-insensitive noise. Depending on the sign of the quantity $(\overline{n}_1 - \overline{n}_2) \cos 2(\theta_0 - \varphi)$ the entanglement between the signal and idler can either decrease or increase as compared to the phase insensitive case. Suppose $\overline{n}_1 > \overline{n}_2$, then the maximum increment for entanglement is achieved if $\varphi = \theta_0$, which corresponds to the narrowest possible phase distribution. In contrast, the  maximum decrement for entanglement takes place if $\varphi = \theta_0 + \pi/2$, which corresponds to the widest possible phase distribution of the pump field, see Fig.~\ref{figure-3}(b).

The behavior of entanglement at a longer timescale ($gt|\alpha_0| > 1$) can be obtained within the generalized parametric approximation. In the case $\overline{n}_1,\overline{n}_2 \ll |\alpha_0|^2$, we get the following formula for coefficients~\eqref{rho-n-m} in the density operator~\eqref{dmatrix-two-mode}:
\begin{eqnarray}
&& \!\!\!\!\!\!\!\!\!\! \varrho_{is}^{nm} =  \frac{(-i)^{n-m} \alpha_0^n (\alpha_0^m)^{\ast} \tanh^{n+m} gt|\alpha_0|}{|\alpha_0|^{n+m} \cosh^2 gt|\alpha_0|}   \nonumber\\
&& \!\!\!\!\!\!\!\!\!\! + \frac{1}{4}(-i)^{n-m} \left( \overline{n}_1 \frac{\partial^2}{\partial [{\rm Re}(\alpha e^{-i \varphi)}]^2} + \overline{n}_2 \frac{\partial^2}{\partial [{\rm Im}(\alpha e^{-i \varphi)}]^2} \right)  \nonumber\\
&& \!\!\!\!\!\!\!\!\!\! \times \left. \frac{\alpha^n (\alpha^m)^{\ast} \tanh^{n+m} gt|\alpha|}{|\alpha|^{n+m} \cosh^2 gt|\alpha|} \right\vert_{\alpha = \alpha_0} + o\left(\frac{\overline{n}_1 + \overline{n}_2}{|\alpha_0|^2}\right). 
\end{eqnarray}

Let us summarize the results of this section. The admixture of the phase-sensitive noise to a coherent pump can either decrease or increase the entanglement between the signal and the idler depending on the relation between the phases of the coherent signal ($\theta_0$) and the noise ($\varphi$), Eq.~\eqref{N-phase-insensitive-noise}. If the dominant noise component is aligned with the direction of coherent state $\ket{\alpha_0}$ in the phase space, then the entanglement reaches its maximum value. In contrast, if the dominant noise component is perpendicular to the direction of coherent state $\ket{\alpha_0}$ in the phase space, then the entanglement is mimimal. This illustrates that the signal-idler entanglement is much more sensitive to a phase distribution of the pump rather than to its amplitude distribution in the phase space.

\section{Effect of dephasing in pump} \label{section-phase}

Dephasing in the fixed pump mode is a stochastic process, which can be caused by several reasons:  phase diffusion in a laser~\cite{henry-1986}, temporal variations of the refractive index in the propagating medium, e.g., the atmosphere and fluctuations of the optical path length~\cite{Semenov2010}. Dephasing of a coherent state $\ket{\alpha_0}$ results in a mixed pump state $\varrho_p = \int f(\varphi) \ket{\alpha_0 e^{i\varphi}}\bra{\alpha_0 e^{i\varphi}} d\varphi$, where $f(\varphi)$ is a phase distribution function. In this section, we consider three kinds of phase distributions: a narrow distribution with the characteristic width $\Delta\theta \ll 1$, a uniform distribution $f(\varphi) = \frac{1}{2\pi}$, $\varphi \in [0,2\pi)$, and a general phase distribution.

\subsection{Small dephasing}

We model the effect of small dephasing on a coherent state $\ket{\alpha_0}$, $\alpha_0 = |\alpha_0|e^{i\theta_0}$, by a Gaussian phase distribution with the standard deviation $\Delta\theta \ll 1$, which leads to the following $P$-function:
\begin{equation}
\label{small-dephasing}
P(|\alpha|e^{i\theta}) = \frac{ \delta(|\alpha| - |\alpha_{0}|) }{|\alpha_{0}| \sqrt{2\pi(\Delta\theta)^2}} \,  \exp\left(-\frac{(\theta-\theta_{0})^2}{2(\Delta\theta)^2} \right),
\end{equation} 

\noindent where $\delta$ is the Dirac delta function. Similar phase smearing is studied in the context of a squeezed coherent pump in Ref.~\cite{HILLERY}. 

In the perturbation theory, we substitute Eq.~\eqref{small-dephasing} into Eq.~\eqref{c-coefficients} and get
\begin{eqnarray}
c_{01} &=& |\alpha_{0}|e^{-(\Delta\theta)^2/2 - i\theta_{0}}, \\ 
c_{11} &=& |\alpha_{0}|^2, \\ 
c_{02} &=& |\alpha_{0}|^2 e^{-2(\Delta\theta)^2 - i 2\theta_{0}}.
\end{eqnarray}

\noindent The linear entropy and negativity of the idler-signal state read as follows in the perturbation theory:
\begin{eqnarray}\label{le-neg-small-dephasing}
&& S_{L} = 2 |\alpha_0|^2 (gt)^2 \left( 1 - e^{-(\Delta\theta)^2} \right) + o(g^2t^2), \\
&& N = gt|\alpha_{0}|e^{-(\Delta\theta)^2 / 2} +(gt|\alpha_{0}|)^2 e^{-2(\Delta\theta)^2} + o(g^2t^2). \qquad
\end{eqnarray}

\begin{figure}
\includegraphics[width=8cm]{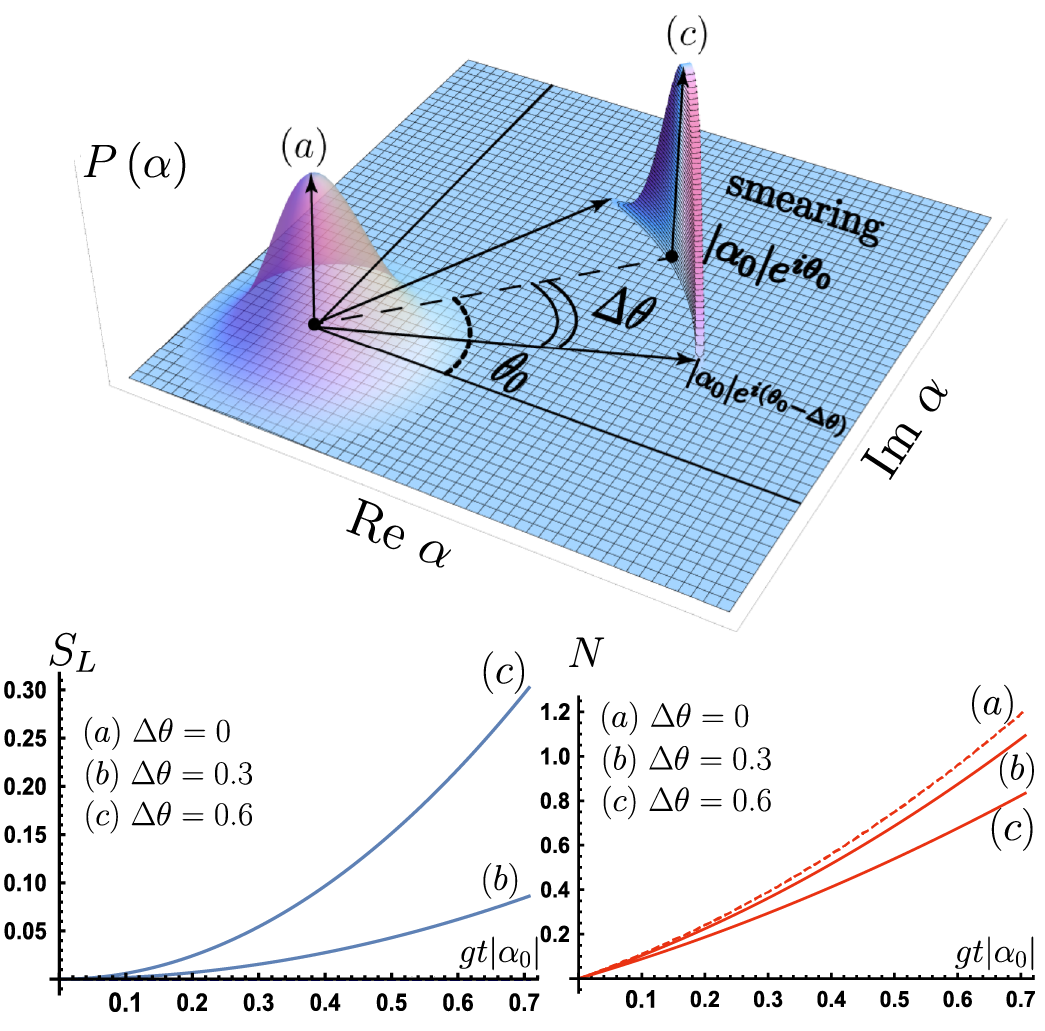}
\caption{Top: the $P$-function of the thermal state (1) and the dephased coherent state (2). Linear entropy (bottom left) and negativity (bottom right) of the signal-idler field for various dephasing parameters $\Delta\theta$: (a) $\Delta\theta = 0$, (b) $\Delta\theta = 0.3$, (c) $\Delta\theta = 0.6$. Linear entropy vanishes in the case (a).}
\label{figure-4}
\end{figure}

Note that the negativity~\eqref{le-neg-displaced-thermal} for the displaced thermal state~\eqref{displaced-thermal} is greater than the negativity of the dephased coherent state~\eqref{small-dephasing} for all $\Delta\theta > 0$. This behavior can be ascribed to the fact that the displaced thermal state is symmetric with respect to $\alpha_0$, i.e. the states $\ket{\alpha_0 + \beta}$ and $\ket{\alpha_0 - \beta}$ equally contribute to the output, whereas the dephased coherent state~\eqref{small-dephasing} is not symmetric with respect to $\alpha_0$ in the phase space, see Fig.~\ref{figure-4}. This observation also implies that the phase distribution of the pump plays a much more important role on the entanglement of idler and signal modes than the amplitude distribution of the pump.

In the generalized parametric approximation, we substitute~\eqref{small-dephasing} in~\eqref{rho-n-m} and get
\begin{eqnarray}\label{dmatrix -dephasing}
&& \varrho_{is}^{nm} = \left( 1-|\lambda(\alpha_{0})|^2 \right) |\lambda(\alpha_{0})|^{n+m} \nonumber\\
&& \times \exp\left[ i(\theta_{0}-\frac{\pi}{2})(n-m) - \frac{(n-m)^2 (\Delta \theta)^2}{2} \right].
\end{eqnarray}

The linear entropy and negativity of the idler-signal field read respectively,
\begin{eqnarray}
 S_L &=& \frac{2}{1+|\lambda(\alpha_0)|^2} \bigg\{ |\lambda(\alpha_0)|^2 - \left[ 1-|\lambda(\alpha_0)|^2 \right] \nonumber\\
&& \qquad\qquad \times  \sum_{k=1}^{\infty} |\lambda(\alpha_0)|^{2k} e^{-k^2 (\Delta \theta)^2} \bigg\}, \\
N &=& \sum_{k=1}^{\infty} |\lambda(\alpha_0)|^k e^{-k^2 (\Delta \theta)^2 / 2}. \label{negativity-gpa-dephasing}
\end{eqnarray}

The formula \eqref{negativity-gpa-dephasing} expresses the negativity in terms of an infinite sum and reduces to Eq.~\eqref{le-neg-small-dephasing} if one considers terms up to the second power of $gt$ only. In general, the infinite sum can be bounded from below by using the inequality $e^{-x} \geq 1 - x$ for real $x$. This leads to the following estimation:
\begin{eqnarray}
N & \geq & \sum_{k=1}^{\infty} |\lambda(\alpha_0)|^k \left[ 1 - \frac{k^2 (\Delta \theta)^2}{2} \right] \nonumber\\
& = & \frac{|\lambda(\alpha_0)|}{1-|\lambda(\alpha_0)|} \left[ 1 - \frac{(1+|\lambda(\alpha_0)|)(\Delta \theta)^2}{2(1-|\lambda(\alpha_0)|)^2} \right] \nonumber\\
& = &  \frac{1}{2}\left( e^{2gt|\alpha_0|} - 1 \right) \left[ 1 -   \left( e^{4gt|\alpha_0|} + e^{2gt|\alpha_0|} \right) \frac{(\Delta \theta)^2}{4} \right]. \nonumber
\end{eqnarray}

Let us summarize the results of this section. Dephasing of the coherent pump diminishes both the purity and the entanglement of the signal-idler field. Deviation of purity and negativity from the correspoinding values for a genuinely coherent pump is quadratic with respect to a dephasing parameter $\Delta\theta$.

\subsection{Phase-averaged pump}

The phase averaged coherent state is defined through
\begin{equation}
\varrho_p = \int_0^{2\pi} \ket{\alpha_0 e^{i\varphi}}\bra{\alpha_0 e^{i\varphi}} \frac{d\varphi}{2\pi}
\end{equation}

\noindent and describes the complete dephasing. It is realized experimentally as an ensemble of coherent states with random phases, with the phase shift being created stochastically by a mirror mounted on a piezoelectric movement~\cite{bondani-2009,allevi-2013}. The $P$-function for such a state reads
\begin{equation} \label{P-pacs}
P(\alpha) = \frac{ \delta(|\alpha| - |\alpha_{0}|) }{2\pi |\alpha_{0}|}.
\end{equation}

In full analogy with the thermal state~\eqref{thermal}, the $P$-function of the phase averaged coherent state is invariant to rotations in the phase space, which leads to vanishing coefficients $c_{nm}$ and $\varrho_{is}^{nm}$ if $n \neq m$. As a result, the negativity vanishes too,
\begin{equation}
N=0.
\end{equation}

In the perturbation theory, we have $c_{11}=|\alpha_0|^2$ and
\begin{equation}\label{le-neg-thermal_new}
S_{L} = 2 |\alpha_0|^2 g^2 t^2  + o(g^2t^2).
\end{equation}

In the generalized parametric approximation, we combine~\eqref{P-pacs}, \eqref{rho-n-m}, and~\eqref{S-gpa} and get
\begin{equation}
S_L = 1 - \frac{1}{\cosh 2gt|\alpha_0|}.
\end{equation}

Comparing the purity of $\varrho_{is}$ for the phase averaged coherent pump and that for the thermal pump of the same intensity [Eq.~\eqref{le-neg-thermal} with $\overline{n} \sim |\alpha_0|^2$], we see that they coincide. Therefore, the thermal pump is indistinguishable from the phase averaged coherent pump from the viewpoint of the idler and signal modes.

Let us summarize the results of this section. Phase-averaged coherent pump cannot generate entanglement between the signal and the idler because it has a uniform phase distribution. Similarly to the case of the thermal pump, only classical correlations are present in the output signal-idler field. The purity of the signal-idler field equals $1/\cosh 2gt|\alpha_0|$ and is less than the purity of the signal-idler field for a genuinely coherent pump (that is equal to $1$).

\subsection{General phase distribution} \label{section-general}

Consider the pump state $\varrho_p = \int_0^{2\pi} f(\varphi) \ket{\alpha_0 e^{i\varphi}}\bra{\alpha_0 e^{i}} d\varphi$. Define $L(\theta) = f ( ( \theta - {\rm arg} \alpha_0 ) {\rm mod} 2\pi )$, then the $P$-function of $\varrho_p$ reads
\begin{equation} \label{P-general-phase}
P(|\alpha|e^{i\theta}) = \frac{\delta(|\alpha|-|\alpha_0|)}{|\alpha_0|} L(\theta),
\end{equation}

\noindent where $L(\theta)$ is a general phase distribution or quasi-distribution such that $\int_0^{2\pi} L(\theta) d\theta = 1$. 

\begin{figure}
\includegraphics[width=8cm]{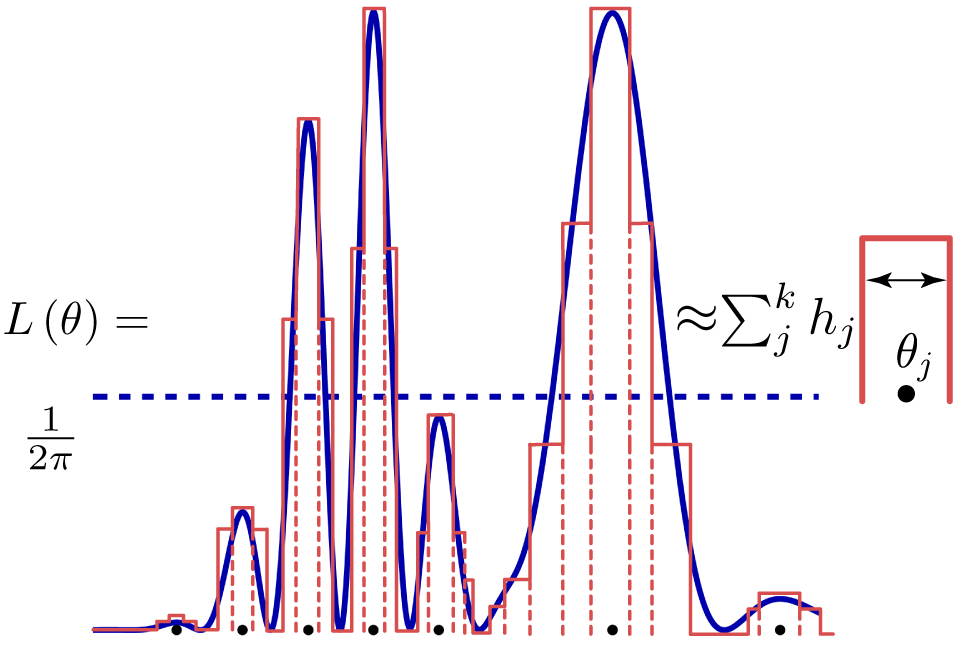}
\caption{Histogram approximation of a general continuous phase distribution function $L(\theta)$. The dashed line illustrates the uniform distribution function $L\left(\theta \right) = 1/2\pi$.}
\label{figure-5}
\end{figure}

Suppose that the function $L(\theta)$ is continuous; then it can be approximated by a histogram function with $k$ bins, see Fig.~\ref{figure-5}. Let $\theta_j$ be a midpoint of $j$th bin and $\Delta\theta_j$ be the size of $j$th bin, then 
\begin{equation}\label{HIST_APR}
L(\theta) \approx \sum_{j=1}^{k} h_j u_j (\theta), 
\end{equation}

\noindent where $u_j(\theta)$ is a rectangular distribution function on interval $[\theta_j - \frac{1}{2}\Delta\theta_j, \theta_j + \frac{1}{2}\Delta\theta_j)$ and coefficients $h_j$ satisfy the normalization condition $\sum_{j=1}^{k} h_{j} = 1$. Combining~\eqref{HIST_APR}, \eqref{P-general-phase}, and~\eqref{c-coefficients}, we get parameters of the down-converted two-mode state in the perturbation theory:
\begin{equation}\label{CMN}
c_{mn} = |\alpha_{0}|^{m+n} \sum_{j=1}^{k} h_{j} \, {\rm{sinc}}\left(\tfrac{1}{2} \Delta\theta_{j}(m-n)\right) e^{i\theta_{j}(m-n)},
\end{equation}

\noindent where ${\rm sinc}(x) = \frac{{\rm sin}x}{x}$. As an example consider a mixture of two narrow uniform distributions with equal weights and the midpoints $\theta'$ and $\theta''$. Since the distributions are narrow, ${\rm{sinc}}\left(\frac{1}{2} \Delta\theta_{j}(m-n)\right) \approx 1$ and $|c_{01}| \approx \left\vert \alpha_{0} \cos \left(\tfrac{1}{2}(\theta'-\theta'')\right) \right\vert$, $|c_{11}| \approx |\alpha_{0}|^2$, $|c_{02}| \approx |\alpha_{0}|^2 |\cos{\left(\theta'-\theta''\right)}|$. The linear entropy and the negativity measure of entanglement for the two-mode idler-signal state read
\begin{eqnarray}
&& \label{le-example} \!\!\!\!\!\!\!\!\! S_{L} \approx 2g^2t^2\sin^2{\tfrac{1}{2}(\theta'-\theta'')},\\
&& \label{n-example}\!\!\!\!\!\!\!\!\! N \approx gt|\alpha_{0}| \left\vert \cos{\tfrac{1}{2}(\theta'-\theta'')} \right\vert +(gt|\alpha_{0}|)^2 \left\vert \cos{\left(\theta'-\theta''\right)} \right\vert. \quad
\end{eqnarray}

\noindent The quantities~\eqref{le-example} and~\eqref{n-example} oscillate with the increase of $\theta'-\theta''$ and resemble the constructive interference if $\theta' - \theta'' = 0$ and destructive interference if $\theta' - \theta'' = \pi$. 

So far in this section we have been considering the factorized $P$-functions of the pump with respect to the amplitude and the phase, i.e. $P(|\alpha|e^{i\theta}) = A(|\alpha|) L(\theta)$. To deal with a general continuous function $P(|\alpha|e^{i\theta})$, we use the Schmidt decomposition
\begin{equation}\label{SCH_D}
P(|\alpha|e^{i\theta}) = \sum_{r} \lambda_r A_r(|\alpha|) L_r(\theta),
\end{equation}

\noindent where the real functions $A_r(|\alpha|)$ and $L_r(\theta)$ satisfy the conditions $\int_{0}^{\infty} A_r(|\alpha|) A_s (|\alpha|) = \delta_{rs}$ and $\int_{0}^{2 \pi} L_r(\theta) L_s(\theta) d\theta = \delta_{rs}$. The functions $L_r(\theta)$ can be further approximated by histogram functions~\eqref{HIST_APR} with possibly negative coefficients $h_j$ such that $\sum h_j = \int_{0}^{2 \pi} L_r(\theta) d\theta$. This approach allows one to deal with general continuous functions $P(\alpha)$ including those describing non-classical states of light~\cite{kiesel-2009,gehrke-2012,ryl-2015}.

In the generalized parameteric approximation, the use of formulas~\eqref{P-general-phase} and \eqref{SCH_D} significantly simplifies the calculation of coefficients~\eqref{rho-n-m} too.

Let us summarize the results of this section. Phase distribution of the pump plays the crucial role in the entanglement between the signal and the idler. If the phase distribution has several peaks, then the entanglement is sensitive to a relative phase difference between the peaks. For a general $P$-function of the pump one can exploit the decomposition~\eqref{SCH_D} over factorized amplitude-phase quasidistributions.

\section{Pump with Kerr squeezing} \label{section-kerr}

The third order nonlinear susceptibility $\chi^{(3)}$ of a medium results in the intensity-dependent refractive index known as the optical Kerr effect. This effect is significant in some glasses~\cite{lenz-2000} and optical fibres~\cite{dziedzic-1981,agrawal-1995}. As a result of the optical Kerr effect, the quantum state of light experiences an intensity-dependent
phase shift leading to a banana-shape quadrature squeezing~\cite{bergman-1991,wilson-gordon-1991,silberhorn-2001}. As the radiation in the pump mode is to be quite intense to observe the nontrivial downconverted state in OPG, the pump itself is subjected to the optical Kerr effect. This happens if the radiation from the pump source (e.g., a laser) is delivered to the downconversion crystal through a fiber. In this section, we study how the Kerr-modulated pump affects the entanglement and purity of the idler and signal modes. 

\begin{figure*}
\centering
\includegraphics[width=\textwidth]{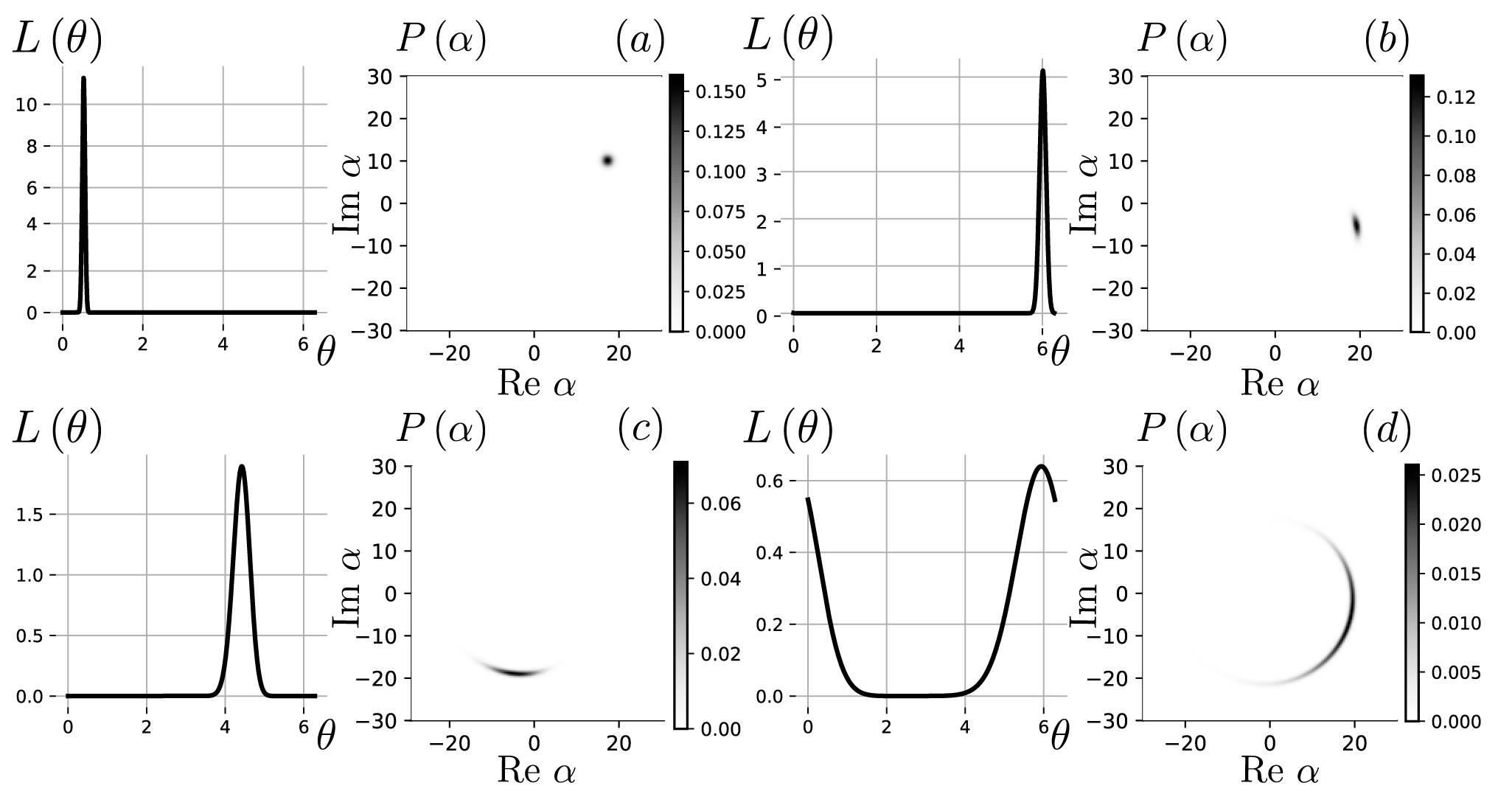}
\caption{Phase distribution $L(\theta)$ and the Glauber-Sudarshan function $P(\alpha)$ of the Kerr-modulated displaced thermal state with parameters $\overline{n} = 1$ and $|\alpha_{0}| = \sqrt{399}$ (the average number of photons is 400) for various Kerr interaction constants $g_K t_K$: (a) $0$, (b) $0.001$, (c) $0.003$, (d) $0.009$.}
\label{figure-6}
\end{figure*}

\begin{figure}
\centering
\includegraphics[width=8.5cm]{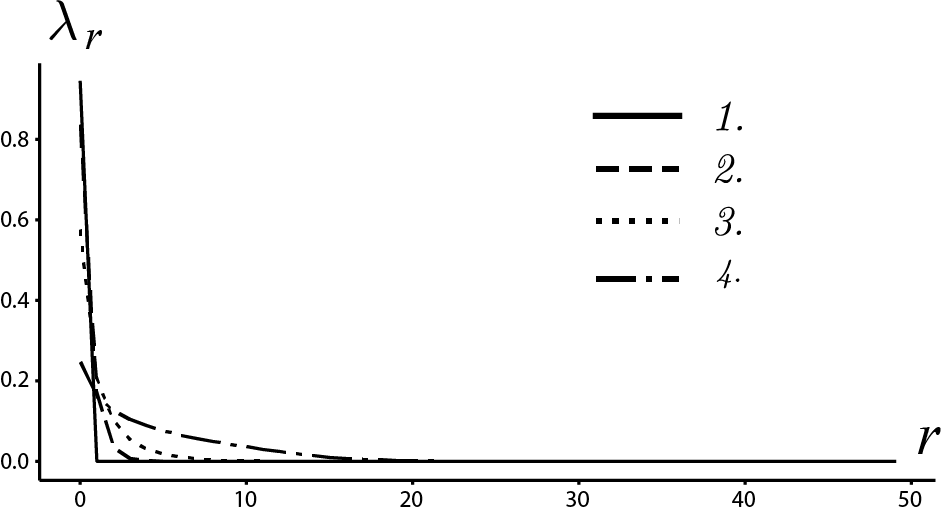}
\caption{Distribution of the Schmidt coefficients $\lambda_r$ in Eq.~\eqref{SCH_D} for the Kerr medium output. The input is a displaced thermal state~\eqref{displaced-thermal} with $|\alpha_{0}| = \sqrt{399}$ and $\overline{n} = 1$ (the average number of photons is 400). Labels $1,2,3,4$ correspond to the Kerr interaction constants $g_{K}t_{K} = 0, 0.001, 0.003, 0.009$, respectively. The decomposition is performed numerically via rasterization of $P$-function in Fig.~\ref{figure-6} in the range ${\rm Re}\,\alpha, {\rm Im}\,\alpha \in [-25; 25]$ with $400$ points in each direction.}
\label{figure-7}
\end{figure}

The optical Kerr effect in the pump mode is described a quartic interaction Hamiltonian $H_{\rm int} = g_{K} (a_p^{\dag})^2 a_p^2 = g_{K} n_p(n_p -1)$, where $n_p = a_p^{\dag }a_p$ and $g_{K} \propto \chi^{(3)}$~\cite{azuma-2008,Hillery2009}. We denote $t_K$ the time of Kerr modulation, then the modified pump state $\widetilde{\varrho}_p$ is related with the original pump state $\varrho_p$ through
\begin{equation} \label{rho-Kerr}
\widetilde{\varrho}_p = \exp\Big[ - i g_K t_K n_p (n_p - 1) \Big] \varrho_p \exp \Big[ i g_K t_K n_p (n_p - 1) \Big]. 
\end{equation}

Clearly, if the original pump state $\varrho_p$ is thermal or phase averaged, then $\widetilde{\varrho}_p = \varrho_p$. The Kerr medium actually modifies only those states $\varrho_p$ that exhibit some degree of coherence. In view of this, further in this section we focus on the coherent state $\varrho_p = \ket{\alpha_0}\bra{\alpha_0}$ and the displaced thermal state~\eqref{displaced-thermal} as inputs to the Kerr medium. 

\begin{figure}
\centering
\includegraphics[width=8.5cm]{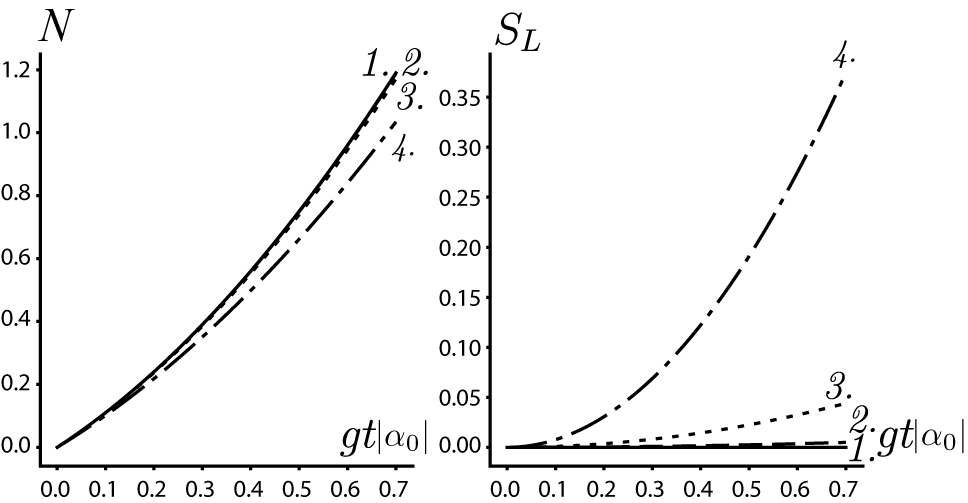}
\caption{
Negativity and linear entropy of the two-mode state of signal and idler generated by a coherent pump subjected to the Kerr effect ($|\alpha_{0}| = 20$). Labels $1,2,3,4$ correspond to the Kerr interaction constants $g_{K}t_{K} = 0, 0.001, 0.003, 0.009$, respectively.}
\label{figure-8}
\end{figure}

\begin{figure*}
\centering
\includegraphics[width=\textwidth]{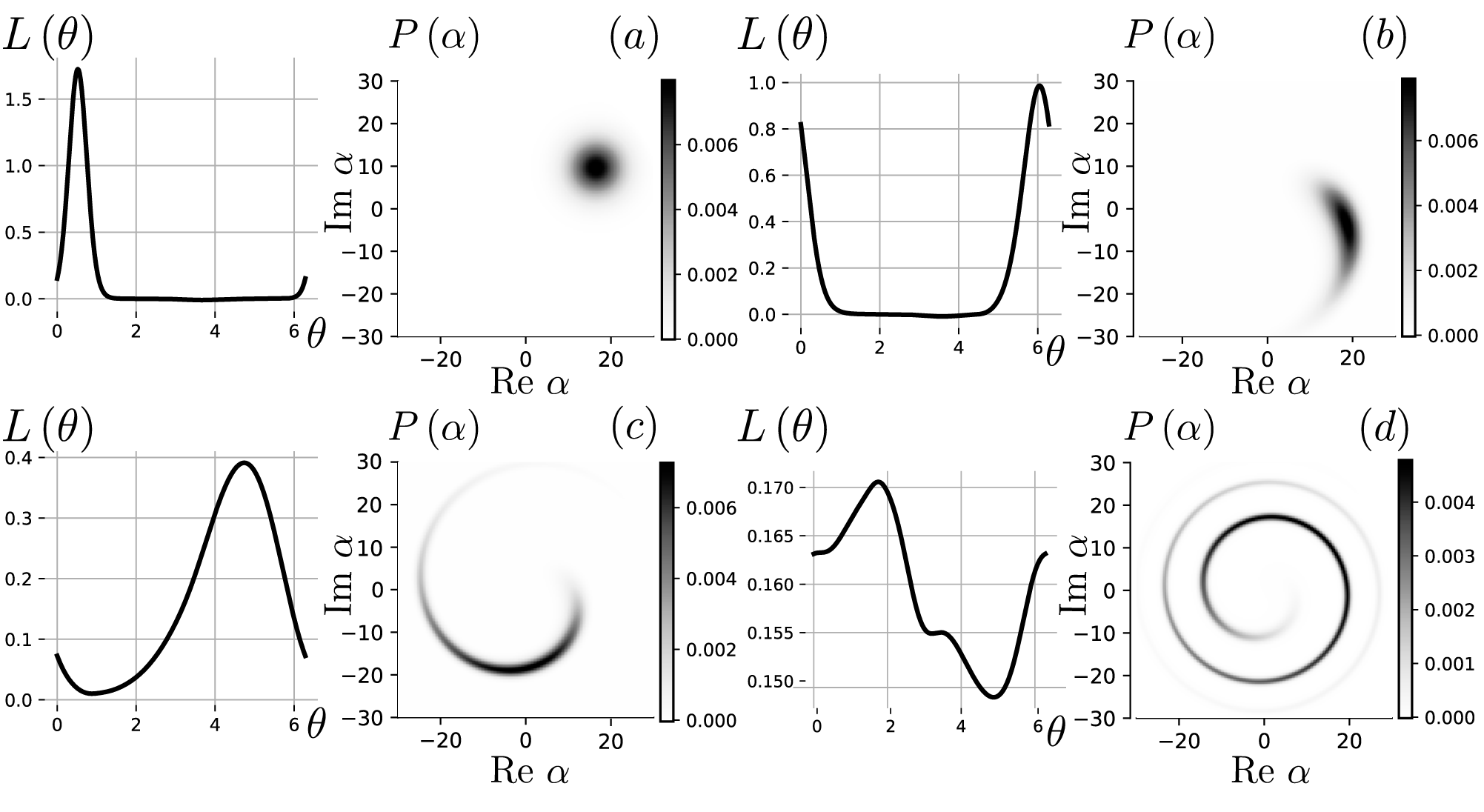}
\caption{Phase distribution $L(\theta)$ and the Glauber-Sudarshan function $P(\alpha)$ of the Kerr-modulated displaced thermal state with parameters $\overline{n} = 39$ and $|\alpha_{0}| = 19$ (the average number of photons is 400) for various Kerr interaction constants $g_K t_K$: (a) $0$, (b) $0.001$, (c) $0.003$, (d) $0.009$.}
\label{figure-9}
\end{figure*}

\begin{figure}
\centering
\includegraphics[width=8.5cm]{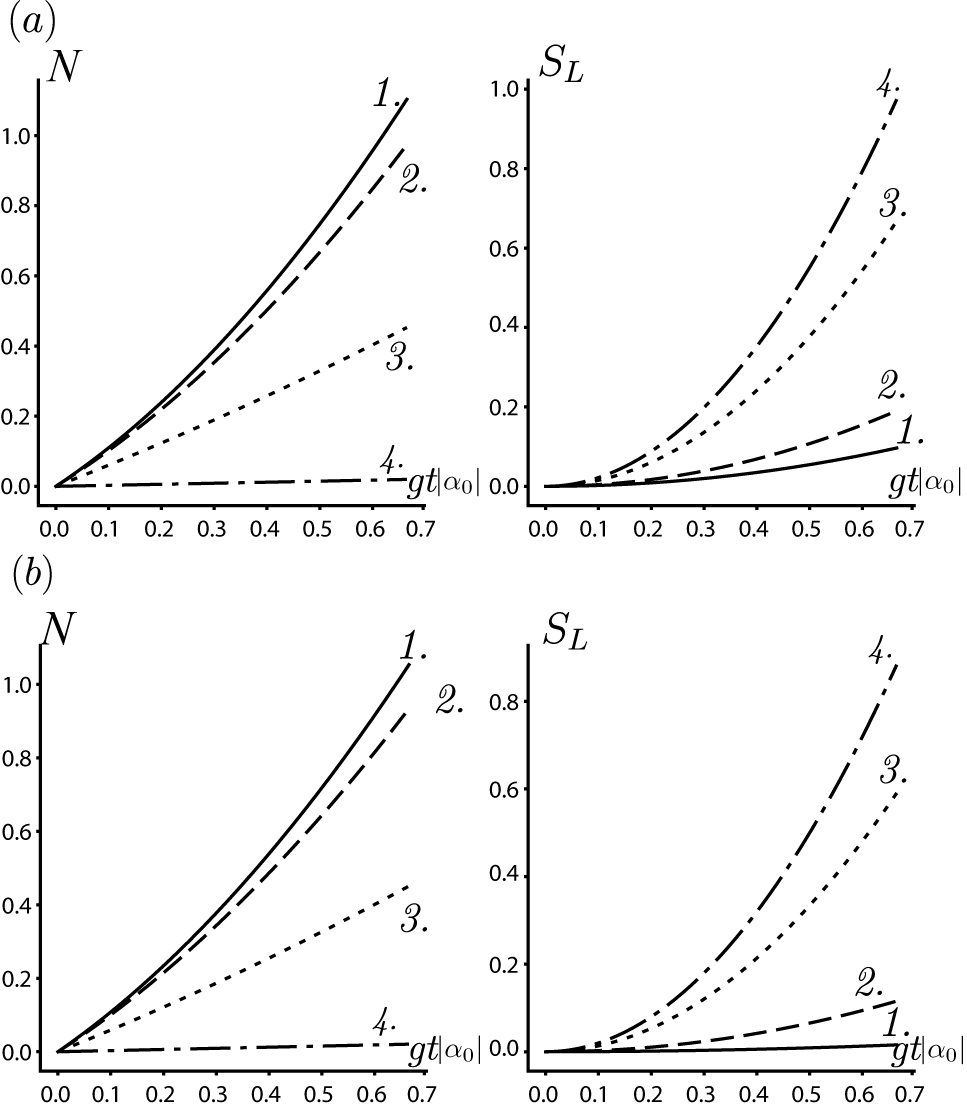}
\caption{Negativity and linear entropy of signal-idler two-mode state obtained in OPG with the Kerr-modulated displaced thermal state as a pump. The displaced thermal state has parameters $\overline{n} = 39$ and $|\alpha_{0}| = 19$ (the average number of photons is 400). Labels $1,2,3,4$ correspond to the Kerr interaction constants $g_{K}t_{K} = 0, 0.001, 0.003, 0.009$, respectively. (a)~Numerical calculations with the use of exact $P$-function~\eqref{P-through-Q}. (b)~Approximate calculation with the use of phase distribution function~\eqref{L-displaced-thermal}.}
\label{figure-10}
\end{figure}

To use formula~\eqref{c-coefficients} we need to find the Glauber-Sudarshan function for the output of Kerr medium. The straightforward approach is to express the $P$-function through the Husimi-Kano function $Q(\xi) = \pi^{-1} \bra{\xi} \widetilde{\varrho}_p \ket{\xi}$ as follows:
\begin{equation} \label{P-through-Q}
P(\alpha) = \frac{1}{(2\pi)^2} \int Q(\xi) \exp \left( \frac{|\xi|^2}{4} + \frac{\xi \alpha^{\ast} + \xi^{\ast} \alpha}{2} \right) d^2\xi. 
\end{equation}

\noindent This approach works well if the input to the Kerr medium is the displaced thermal state~\eqref{displaced-thermal} with $\overline{n} > 0$. In Fig.~\ref{figure-6}, we depict the $P$-function of the Kerr medium output for different times $t_K$ of the Kerr modulation. However, there are two complications related with such an approach: (i) the use of Eq.~\eqref{P-through-Q} conceals the physical properties of $P$-function and the effect of parameters $\alpha_0$, $\overline{n}$, $g_K t_K$ on the purity and entanglement of OPG two-mode states; (ii) Eq.~\eqref{P-through-Q} is not applicable directly to the coherent input $\ket{\alpha_0}\bra{\alpha_0}$ to the Kerr medium because of the singular character of the $P$-function~\cite{Walls,gehrke-2012}.

Due to the above complications we develop an approximate treatment of the $P$-function based on the factorization
\begin{equation} \label{factorization}
P(|\alpha|e^{i\theta}) \approx A(|\alpha|) L(\theta),
\end{equation}
\noindent where $L(\theta)$ is the phase distribution function defined through
\begin{equation} \label{L-definition}
L(\theta) = \int_0^{\infty} P(|\alpha|e^{i\theta}) |\alpha| d|\alpha|.
\end{equation}

\noindent We verify the validity of approximation~\eqref{factorization} for the Kerr-modulated displaced thermal state by numerically calculating the Schmidt coefficients $\lambda_r$ in formula~\eqref{SCH_D} for such a state. In Fig.~\ref{figure-7} we depict the distribution of Schmidt coefficients for different times $t_K$. The approximation~\eqref{factorization} is valid if the coefficients $\lambda_r$ decay rapidly with $r$. This takes takes place indeed if $\Delta\theta \ll 2\pi$, where $\Delta\theta$ is the characteristic width of the phase distribution function $L(\theta)$.

Let us estimate the characteristic phase spreading of the $P$-function in phase space for the Kerr-modulated pump. The operator $\exp \left( -i a^{\dag}a \varphi \right)$ rotates the phase space by the angle $\varphi$ and transforms the Fock state $\ket{n}$ into $e^{-i n \varphi}\ket{n}$. The operator $\exp \left(-i (a^{\dag})^2 a^2 \varphi \right)$ transforms the Fock state $\ket{n}$ into $e^{-i n(n-1) \varphi}\ket{n}$, which can be interpreted as a rotation in the phase space by the angle $(n-1)\varphi$. It means that the rotation angle for Fock states is linearly proportional to their energy. Therefore, the characteristic phase spreading $\Delta\theta$ of the $P$-function due to the Kerr medium can be estimated through the dispersion of photon numbers $n$ in the Fock states significantly contributing to $\varrho_p$, namely, $\Delta\theta \sim g_K t_K \sqrt{ \ave{n^2} - \ave{n}^2 }$. For the coherent state $\varrho_p = \ket{\alpha_0}\bra{\alpha_0}$ we have
\begin{equation} \label{spreading-coherent}
\Delta\theta_{\rm coh.} \sim g_K t_K |\alpha_0|.
\end{equation}

\noindent In contrast to the coherent state, the displaced thermal state~\eqref{displaced-thermal} itself has the phase dispersion $\overline{n}/|\alpha_0|^2$ if $|\alpha_0|^2 \gg \overline{n}$. This dispersion is enhanced by dispersion $(g_K t_K)^2 ( \ave{n^2} - \ave{n}^2 )$ when the pump light passes through the Kerr medium. We use the known photon number distribution function~\cite{adam-1995} and get
\begin{equation} \label{standard-deviation-dis-th}
\Delta\theta_{\rm dis.th.} \sim  \sqrt{\frac{\overline{n}}{|\alpha_0|^2} + (g_K t_K)^2 |\alpha_0|^2 ( 2\overline{n} + 1 )}
\end{equation}

\noindent provided $|\alpha_0|^2 \gg \overline{n}$. 

We focus on experimentally achievable modulation constants $g_K t_K \ll 1$ such that the characteristic phase spreading $\Delta\theta \ll 2\pi$. If this is the case, then the Glauber-Sudarshan function $P(|\alpha|e^{i\theta})$ of $\widetilde{\varrho}_p$ is well approximated by a factorized form $A(|\alpha|) L(\theta)$.

The amplitude distributions $A_{\rm coh.}(|\alpha|)$ and $A_{\rm dis.coh.}(|\alpha|)$ for the initially coherent state $\varrho_p = \ket{\alpha_0}\bra{\alpha_0}$ and the displaced thermal state~\eqref{displaced-thermal}, respectively, read
\begin{eqnarray}
&& A_{\rm coh.}(|\alpha|) = \frac{1}{|\alpha_0|} \delta(|\alpha| - |\alpha_0|), \\
&& A_{\rm dis.th.}(|\alpha|) = \frac{1}{\sqrt{\pi \overline{n}} |\alpha_0|} \exp\left( -\frac{(|\alpha| - |\alpha_0|)^2}{\overline{n}} \right).
\end{eqnarray}

\noindent and satisfy $\int_0^{\infty} A_{\rm coh.}(|\alpha|) |\alpha| d|\alpha| = 1$ and $\int_0^{\infty} A_{\rm dis.th.}(|\alpha|) |\alpha| d|\alpha| \approx 1$ if $|\alpha_0|^2 \gg \overline{n}$. 

Substituting the approximation $P(|\alpha|e^{i\theta}) \approx A(|\alpha|) L(\theta)$ in Eq.~\eqref{c-coefficients}, we get
\begin{equation} \label{C-angle}
c_{mn} \approx |\alpha_0|^{m+n} \int_{0}^{2\pi}  L(\theta) e^{i\theta(m-n)} d\theta.
\end{equation}

\noindent It is the phase distribution $L(\theta)$ that significantly contributes to the coefficients $c_{mn}$ and defines the purity and the entanglement of signal and idler modes. 

In Appendix~\ref{appendix-a}, we analytically calculate the phase distribution $L_{\rm coh.}(\theta)$ in the case of the coherent state $\varrho_p = \ket{|\alpha_0|e^{i\theta_0}}\bra{|\alpha_0|e^{i\theta_0}}$ as an input to the Kerr medium; the result is
\begin{equation} \label{L-coherent}
L_{\rm coh.}(\theta) = e^{-|\alpha_0|^2} \sum_{k=0}^{\infty} \frac{|\alpha_0|^{2k}}{4\pi k!} \, D_{k}\Big( \theta - \theta_0 + g_K t_K (2k-1) \Big),
\end{equation}

\noindent where $D_k(x) = \frac{\sin(k+1/2)x}{\sin(x/2)}$ is the Dirichlet kernel. We further substitute Eq.~\eqref{L-coherent} into Eq.~\eqref{C-angle} and numerically calculate the linear entropy and negativity of the signal and idler modes. The results are presented in Fig.~\ref{figure-8}. The greater the Kerr modulation $g_{K}t_{K}$, the less entangled and more mixed the state of signal and idler modes is.

In Appendix~\ref{appendix-b}, we analytically calculate the phase distribution $L_{\rm dis.th.}(\theta)$ in the case of the displaced thermal state~\eqref{displaced-thermal} as an input to the Kerr medium; the result is
\begin{eqnarray} \label{L-displaced-thermal}
&& \!\!\!\!\! L_{\rm dis.th.}(\theta) = \frac{e^{-|\alpha_0|^2/\overline{n}}}{4\pi \overline{n}} \sum_{\scriptsize \begin{array}{c}
k,l = 0,1,2,\ldots  \\
k+l \text{~is~even}
\end{array}} \frac{\Gamma \left( \frac{3k+l+4}{4} \right)}{\left(\frac{k+l}{2} \right)!\left(\frac{k-l}{2} \right)!} \nonumber\\
&& \!\!\!\!\! \times \left( \tfrac{|\alpha_0|^2 \overline{n}}{\overline{n}+1}  \right)^{\frac{k-l}{4}}  \left( \tfrac{\overline{n}}{\overline{n}+1} \right)^{\frac{k+l+2}{2}}  {}_1F_1 \left( \tfrac{3k+l+4}{4}; \tfrac{k-l+2}{2}; \tfrac{|\alpha_0|^2}{\overline{n}(\overline{n}+1)} \right) \nonumber\\ 
&& \!\!\!\!\! \times \exp \left\{ -i [\theta - \theta_0 + g_K t_K (k+l-1)] \frac{k-l}{2} \right\},
\end{eqnarray}

\noindent where $\Gamma(x)$ is the gamma function and ${}_1F_1(a;b;z)$ is the confluent hypergeometric function. 

If $\Delta\theta_{\rm dis.th.} \ll 2\pi$, then $L_{\rm dis.th.}(\theta)$ is well approximated by a Gaussian distribution with the standard deviation~\eqref{standard-deviation-dis-th} and one can use the results of section~\ref{section-phase}. If $\Delta\theta_{\rm dis.th.} \sim \pi$, then the Gaussian approximation for $L_{\rm dis.th.}(\theta)$ is not valid, see Fig.~\ref{figure-9}. However, substituting Eq.~\eqref{L-displaced-thermal} into Eq.~\eqref{C-angle}, we still get the coefficients $c_{nm}$ and calculate the linear entropy and negativity of the signal and idler modes, see Fig.~\ref{figure-10}(b). Even in this case, our results are in good agreement with the exact numerical calculation exploiting formula~\eqref{P-through-Q}, cf. Fig.~\ref{figure-10}(a).

Comparing the results for the Kerr-modulated displaced thermal pump (Fig.~\ref{figure-10}) and the Kerr-modulated coherent pump (Fig.~\ref{figure-8}) with the same number of photons, we conclude that the former one leads to a worse entanglement and a lower purity of the signal-idler modes than the latter one. The greater the Kerr modulation $g_K t_K$, the more significant is the advantage of the coherent state.

Let us summarize the results of this section. We have found the phase distributions for a coherent pump subjected to the optical Kerr effect, Eq.~\eqref{L-coherent}, and a noisy coherent pump subjected to the optical Kerr effect, Eq.~\eqref{L-displaced-thermal}. Characteristic widths of such distributions are given by formulas~\eqref{spreading-coherent} and~\eqref{standard-deviation-dis-th}, respectively. Using the obtained phase distributions and Eq.~\eqref{C-angle}, we have managed to calculate the negativity and the linear entropy for the signal-idler fields generated with corresponding Kerr modulated pumps.

\section{Effect of incoherent pump on quadratures} \label{section-quadratures}

Homodyne measurements enable one to get access to correlations between the idler and signal fields in terms of quadratures. The crucial fact is that the dispersions of the combined quadratures
\begin{eqnarray}\label{quad}
&& X_- = \frac{X_i - X_s}{\sqrt{2}} = \frac{ a_{i} + a_{i}^{\dagger} -  a_{s} - a_{s}^{\dagger} }{2}, \\
&& P_+ = \frac{P_i + P_s}{\sqrt{2}} = \frac{ a_{i} - a_{i}^{\dagger} + a_{s} - a_{s}^{\dagger} }{2i}
\end{eqnarray}

\noindent can both be significantly below the dispersion of the vacuum field (equal to $\frac{1}{2}$), which is an indication of nonclassical correlations. The experiments deal with quantity $-10 \log_{10}[2\langle(\Delta X_-)^2\rangle]$ that quantifies the squeezing in dB and serves as a quality factor for the prepared entangled state~~\cite{wang-2010,yan-2012,zhou-2015,yu-2016}.

For the idler-signal density operator $\varrho_{is} = \sum_{n,m=0}^{\infty} \varrho_{is}^{nm} \ket{n_i n_s} \bra{m_i m_s}$ we have
\begin{equation} \label{squeezing-through-rho-n-m}
\braket{2 (\Delta X_{-})^2} = \braket{2 (\Delta P_{+})^2} = 1 + 2 \sum_{n=0}^{\infty} n \left( \varrho_{is}^{nn} - {\rm Re} \varrho_{is}^{n,n-1} \right).
\end{equation}

In the case of the coherent pump $\varrho_p = \ket{\alpha_0}\bra{\alpha_0}$ with $\alpha_0 = i|\alpha_0|$, we get the following expression in the parametric approximation~\eqref{rho-is-alpha-gpa}:
\begin{equation}\label{qvar-coh}
\braket{2 (\Delta X_{-})^2} = e^{- 2 gt|\alpha_{0}|}.
\end{equation}

For the incoherent pump the quantity $\braket{2 (\Delta X_{-})^2}$ differs from $e^{- 2 gt|\alpha_{0}|}$. Namely, for the noisy coherent state~\eqref{displaced-thermal} with $\overline{n} \ll |\alpha_0|^2$ we combine~\eqref{rho-n-m-gpa-displaced-thermal} and \eqref{squeezing-through-rho-n-m}, take into account that
\begin{eqnarray}
&& {\rm Re} \left( \frac{\partial^2}{\partial ({\rm Re}\alpha)^2} + \frac{\partial^2}{\partial ({\rm Im}\alpha)^2} \right) \left. \frac{\alpha}{|\alpha|} f(|\alpha|) \right\vert_{\alpha = |\alpha|} \nonumber\\
&& = - \frac{1}{|\alpha|^2} f(|\alpha|) + \left( \frac{\partial^2}{\partial ({\rm Re}\alpha)^2} + \frac{\partial^2}{\partial ({\rm Im}\alpha)^2} \right) f(|\alpha|), \qquad
\end{eqnarray}

\noindent and finally get
\begin{eqnarray}\label{qvar-disp}
&& \braket{2 (\Delta X_{-})^2} = e^{-2gt|\alpha_0|} + \frac{\overline{n}}{4|\alpha_0|^2} \sinh 2gt|\alpha_0| \nonumber\\
&&  +  \frac{\overline{n}}{4} \left( \frac{\partial^2}{\partial ({\rm Re}\alpha)^2} + \frac{\partial^2}{\partial ({\rm Im}\alpha)^2} \right) \left. e^{-2gt|\alpha|} \right\vert_{\alpha = \alpha_0} = e^{-2gt|\alpha_0|} \nonumber\\
&&  + \frac{\overline{n}}{4|\alpha_0|^2} \left[ \sinh 2gt|\alpha_0| + 2gt|\alpha_0|(2gt|\alpha_0| - 1) e^{-2gt|\alpha_0|} \right]. \nonumber\\
\end{eqnarray}

For the dephased coherent state~\eqref{small-dephasing} we combine~\eqref{dmatrix -dephasing} and \eqref{squeezing-through-rho-n-m} and get
\begin{equation} \label{quad-dephasing}
\braket{2 (\Delta X_{-})^2} = e^{-2gt|\alpha_0|} + \left( 1 - e^{-(\Delta \theta)^2 / 2}\right) \sinh 2gt|\alpha_0|.
\end{equation}

\begin{figure}
\centering
\includegraphics[width=8cm]{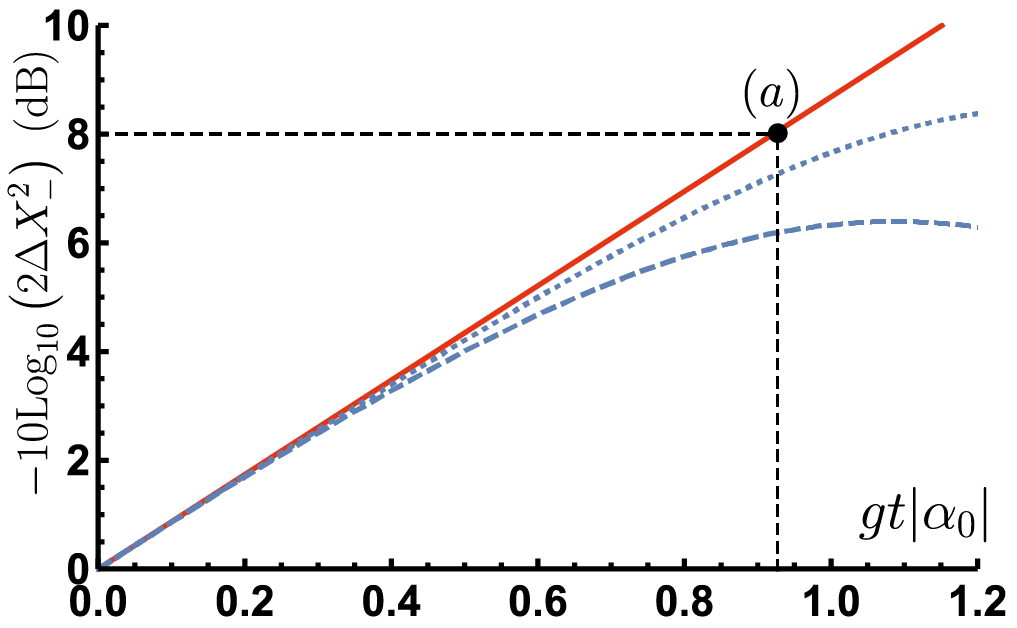}
\caption{Quadrature variance $\braket{2(\Delta X_-)^2}$ in $\rm{dB}$ vs. dimensionless parameter $gt|\alpha_0|$ for different pumps: coherent pump (red solid line), displaced thermal pump (blue dashed line, $\overline{n}/|\alpha_{0}|^2 = 0.1$), and dephased pump (blue dotted line, $\Delta\theta = 0.1$). Point $(a)$ indicates the experimentally achievable value of quadrature squeezing $8$ dB~\cite{zhou-2015}. }
\label{figure-11}
\end{figure}

Fig.~\ref{figure-11} depicts the effect of noisy coherent pump and the dephased coherent pump on the relative quadrature variance $\braket{2(\Delta X_-)^2}$.

Let us summarize the results of this section. We have considered another quantifier of the signal-idler entanglement, namely, the combined quadrature squeezing. We have shown that the squeezing parameter diminishes if some noise is added to a coherent pump. The admixture of the phase-insensitive Gaussian noise is described by Eq.~\eqref{qvar-disp}. The effect of pump dephasing is described by Eq.~\eqref{quad-dephasing}.

\section{Summary and conclusions} \label{section-conclusions}

Current technology provides the sources of intense light beyond conventional lasers, for instance, laser diodes which could be used as a pump in the OPG. The down-converted photons pairs are still created simultaneously with such an incoherent pump, however, they rather correspond to a mixed state $\varrho_{is}$ of the signal and idler modes than to a pure one. In fact, the down-converted modes (signal and idler ones) inherit the incoherent properties of the pump and exhibit less degree of entanglement and purity as compared to the OPG with the coherent pump. These results can also be extended to the field of microwave quantum state engineering with Josephson traveling-wave parametric amplifiers, where the
thermal noise in the pump is inevitable~\cite{zorin-2016,grimsmo-2017}.

We have demonstrated, that the thermal pump and the phase-averaged coherent pump result in no entanglement between the idler and signal modes, the classical correlations are present only. The noisy coherent (displaced thermal) pump and the partially dephased coherent pump produce some entanglement between the idler and signal modes; the degree of entanglement decreases with the growth of thermal contribution and dephasing, respectively. The variance of the combined quadratures exhibits the same behavior. The feature of the phase-sensitive Gaussian noise is that the degree of signal-idler entanglement can either decrease or increase depending on the relation between the phase of the dominant noise component and the pump phase.

We have developed the approach to deal with an arbitrary pump defined by the Glauber-Sudarshan function $P(\alpha)$. The entanglement and the purity of $\varrho_{is}$ are much more sensitive to the phase distribution $L(\theta)$ of the pump rather than to the amplitude distribution $A(|\alpha|)$. The general observation is that the wider the phase distribution $L(\theta)$, the less entangled are the signal and idler modes. As an instructive example, we have analytically calculated the phase distribution $L(\theta)$ for the Kerr-modulated coherent pump and the Kerr-modulated displaced thermal pump, and used this distribution to infer the entanglement properties of the idler and signal modes. 

\begin{acknowledgements}
S.V.V. thanks the Foundation for the Advancement of Theoretical Physics and Mathematics “BASIS” for support under Grant No.17-15-603-1. S.N.F. was partially supported by Program No.0066-2019-0005 of the Ministry of Science and Higher Education of Russia for Valiev Institute of Physics and Technology of RAS. S.N.F. thanks the Russian Foundation for Basic Research for partial support under Project No. 17-07-00994.
\end{acknowledgements}
\onecolumngrid
\appendix
\section{Phase distribution for the Kerr-modulated coherent state} \label{appendix-a}

In this section, we derive the phase distribution $L\left(\theta\right)$ for coherent pump modulated by the Kerr medium. 

We start with a general transformation of the $P$-function in the Kerr medium, which follows from Eq.~\eqref{rho-Kerr} and Ref.~\cite{Sudarshan}:

\begin{eqnarray}\label{P_EVOLUTION}
&& \!\!\!\!\! P_{\rm out}\left(|\alpha|e^{i\theta} \right) 
= \sum_{k,l=0}^{\infty} \frac{\exp\left\{-i\left [\theta+g_{K}t_{K}\left(k+l-1\right)\right] \left(k-l\right) \right\}}{2\pi(k+l)!}    \int\limits_0^{\infty} d\lvert\beta\rvert \int\limits_0^{2\pi} d\varphi \, e^{-\lvert\beta\rvert^2} P_{\rm in}\left(|\beta|e^{i\varphi}\right)\lvert\beta\rvert^{k+l+1} e^{i\varphi\left(k-l\right)} \nonumber\\
&& \times \frac{e^{\lvert\alpha\rvert^2}}{\lvert\alpha\rvert}\left(-\frac{\partial}{\partial\lvert\alpha\rvert}\right)^{k+l}\delta\left(|\alpha|\right),
\end{eqnarray}
\noindent where $P_{\text{in~(out)}}$ is the $P$-function of the Kerr medium input (output).

To find the phase distribution~\eqref{L-definition}, we utilize several mathematical transformations. The idea of the first transformation is based on the extension of the integration limits in Eq.~\eqref{L-definition} and formally include negative values, i.e., $|\alpha| \rightarrow a \in (-\infty,+\infty)$. Denote
\begin{eqnarray}
I_{n}^{(+)} &=& \int_{0}^{\infty} e^{a^2}\left(-\frac{\partial}{\partial a}\right)^{n} \delta(a) da, \\
I_{n}^{(-)} &=& \int_{-\infty}^{0}  e^{a^2}\left(-\frac{\partial}{\partial a}\right)^{n} \delta(a) da = (-1)^n I_n^{(+)}, \quad
\end{eqnarray}

\noindent then the integration of $P_{\rm out}(a e^{i\theta})$ over $a \in (-\infty,+\infty)$ yields
\begin{eqnarray}\label{INTEGRATION2}
&&\int_{-\infty}^{\infty} P_{\rm out}(a e^{i\theta}) a \, da 
= \int_{-\infty}^{0} P_{\rm out}(a e^{i\theta}) a \, da  + \int_{0}^{\infty} P_{\rm out}(a e^{i\theta}) a \, da  = \sum_{k,l=0}^{\infty} (\cdots) \left[ I_{k+l}^{(+)} + I_{k+l}^{(-)} \right] = \nonumber \\
&& = \sum_{k,l=0}^{\infty} (\cdots) \left[ (-1)^{k+l} I_{k+l}^{(+)} + I_{k+l}^{(+)} \right], 
\end{eqnarray}
\noindent where $(\cdots)$ denotes the terms in the first line of Eq.~\eqref{P_EVOLUTION}. Since $(-1)^{k+l} = e^{- i \pi (k-l)}$, we combine $(\cdots)$ and $e^{- i \pi (k-l)}$ and get 
\begin{eqnarray}\label{INTEGRATION3}
&&\int_{-\infty}^{\infty} P_{\rm out}(a e^{i\theta}) a \, da = \int_{0}^{\infty} P_{\rm out}(a e^{i(\theta+\pi)}) a \, da  + \int_{0}^{\infty} P_{\rm out}(a e^{i\theta}) a \, da = L\left(\theta+\pi\right) + L\left(\theta\right).
\end{eqnarray}

On the other hand,
\begin{eqnarray}\label{INTEGRATION}
I_{n}^{(+)} + I_{n}^{(-)}= \int_{-\infty}^{\infty} e^{a^2} \left(-\frac{\partial}{\partial a}\right)^{n} \delta\left(a\right)d a  
=\lvert H_{n}\left(0\right)\rvert = \left\{ \begin{array}{cc}
  0  & \text{if~} n \text{~is odd}, \\
  \frac{n!}{( n/2 )!}   & \text{if~} n \text{~is even},
\end{array} \right.
\end{eqnarray}
\noindent where $H_{n}\left(x\right)$ is the $n$th-order Hermite polynomial. 

Eqs.~\eqref{P_EVOLUTION}, \eqref{INTEGRATION3} and~\eqref{INTEGRATION} imply that 
\begin{eqnarray} \label{L-theta-pi-L-theta}
&& L\left(\theta+\pi\right) + L\left(\theta\right) = \sum_{\scriptsize \begin{array}{c}
k,l = 0,1,2,\ldots  \\
k+l \text{~is~even}
\end{array}} (\cdots) \, \frac{(k+l)!}{\left(\frac{k+l}{2}\right)!} = \sum_{\scriptsize \begin{array}{c}
k,l = 0,1,2,\ldots  \\
k+l \text{~is~even}
\end{array}} \int\limits_0^{\infty} d\lvert\beta\rvert \int\limits_0^{2\pi} d\varphi \, e^{-\lvert\beta\rvert^2} P_{\rm in}\left(|\beta|e^{i\varphi}\right)\lvert\beta\rvert^{k+l+1} \nonumber\\
&& \times \frac{\exp\left\{-i\left [\theta - \varphi +g_{K}t_{K}\left(k+l-1\right)\right] \left(k-l\right) \right\}}{2\pi\left(\frac{k+l}{2}\right)!}.
\end{eqnarray}

Note that the relation 
\begin{equation}
\sum_{n \text{~is~even}} e^{-i \beta n} = \frac{1}{2} \sum_{n \text{~is~even}} \left( e^{- i (\beta+\pi) n /2} + e^{- i \beta n /2} \right)
\end{equation}

\noindent holds for any real $\beta$, so the function
\begin{eqnarray} \label{L-general}
&& \!\!\!\!\! L(\theta) = \sum_{\scriptsize \begin{array}{c}
k,l = 0,1,2,\ldots  \\
k+l \text{~is~even}
\end{array}} \int\limits_0^{\infty} d\lvert\beta\rvert \int\limits_0^{2\pi} d\varphi \, e^{-\lvert\beta\rvert^2} \lvert\beta\rvert^{k+l+1} P_{\rm in}\left(|\beta|e^{i\varphi}\right) \frac{\exp\left\{-i\left [\theta - \varphi +g_{K}t_{K}\left(k+l-1\right)\right] \left(k-l\right) / 2 \right\}}{4\pi\left(\frac{k+l}{2}\right)!} \nonumber\\
\end{eqnarray}
\noindent satisfies Eq.~\eqref{L-theta-pi-L-theta} and is the actual phase distribution of the output of the Kerr medium. 

If the input to the Kerr medium is a coherent state $\ket{\alpha_0}\bra{\alpha_0}$ with $\alpha_0 = |\alpha_0|e^{i\theta_0}$, then $P_{\rm in}\left(|\beta|e^{i\varphi}\right) = |\alpha_0|^{-1} \delta(\beta - |\alpha_0|) \delta(\varphi - \theta_0)$ and Eq.~\eqref{L-general} yields
\begin{eqnarray}
L_{\rm coh.}(\theta) = e^{-|\alpha_0|^2} \sum_{\scriptsize \begin{array}{c}
k,l = 0,1,2,\ldots  \\
k+l \text{~is~even}
\end{array}} \frac{|\alpha_0|^{k+l} }{4\pi\left(\frac{k+l}{2}\right)!}   \times \exp\left\{ -i\left [\theta - \theta_0 +g_{K}t_{K}\left(k+l-1\right)\right] \left(k-l\right) / 2 \right\}. \qquad
\end{eqnarray}

\noindent We make a change of variables, which takes into account that both $k+l$ and $k-l$ are even:
\begin{equation}\label{CHANGE_INDEX}
k + l = 2 x, \quad k - l = 2y, \quad |y| \leq x, \quad x,y = 0,1,2,\ldots
\end{equation}

\noindent Finally, we have 
\begin{eqnarray}
L_{\rm coh.}(\theta) = e^{-|\alpha_0|^2} \sum_{x=0}^{\infty} \sum_{y=-x}^{x} \frac{|\alpha_0|^{2x} e^{-i\left [\theta - \theta_0 +g_{K}t_{K}\left(2x-1\right)\right] y }}{4\pi x!}= e^{-|\alpha_0|^2} \sum_{x=0}^{\infty} \frac{|\alpha_0|^{2x}}{4\pi x!} \, D_{x}\Big( \theta - \theta_0 + g_K t_K (2x-1) \Big), \quad
\end{eqnarray}

\noindent where $D_x(z) = \frac{\sin(x+1/2)z}{\sin(z/2)}$ is the Dirichlet kernel.

\section{Phase distribution for the Kerr-modulated displaced thermal state} \label{appendix-b}

In this section, we find the phase distribution $L(\theta)$ for the output of the Kerr medium, when the input is a displaced thermal state with the $P$-function 
\begin{eqnarray} \label{P-displaced-thermal-app-b}
P_{\rm in}\left(|\beta| e^{i\varphi} \right) = \frac{1}{\pi \overline{n}} \exp \left(- \frac{\left\vert |\beta| e^{i\varphi} - |\alpha_0|e^{i\theta_0}\right\vert^2 }{\overline{n}} \right) = \frac{1}{\pi \overline{n}} \exp\left(- \frac{\lvert\beta \rvert^2}{\overline{n}} - \frac{\lvert\alpha_{0}\rvert^2}{\overline{n}} + \frac{2\lvert\beta\rvert \lvert\alpha_{0}\rvert \cos{\left(\varphi - \theta_{0}\right)}}{\overline{n}}\right). \qquad
\end{eqnarray}

While substituting~\eqref{P-displaced-thermal-app-b} in the general formula~\eqref{L-general}, we integrate over $\varphi$ first and then integrate over $|\beta|$. The first integration yields
\begin{eqnarray}\label{AUX_INTEGRATION_1}
\int_{0}^{2\pi} \exp\left( \frac{2\lvert\beta\rvert\lvert \beta_{0}\rvert \cos{\left(\theta - \theta_{0}\right)}}{\overline{n}}  + i\frac{\varphi\left(k-l\right)}{2} d\varphi \right)= 2\pi e^{i\theta_{0}(k-l)/2} I_{|k-l|/2} \left(\frac{2\lvert\beta\rvert\lvert \alpha_{0}\rvert}{\overline{n}}\right),
\end{eqnarray}

\noindent where $I_{n}(x)$ is the modified Bessel function of the first kind. The second integration yields~\cite{PRUDNIKOV}:
\begin{eqnarray}\label{AUX_INTEGRATION_2}
\int_{0}^{\infty} \exp\left({-\frac{\overline{n}+1}{\overline{n}}} |\beta|^2 \right) \lvert\beta\rvert^{k+l+1} I_{|k-l|/2} \left(\frac{2\lvert\beta\rvert\lvert \alpha_{0}\rvert}{\overline{n}}\right) d\lvert\beta\rvert = \left(\frac{\overline{n}}{2\lvert\alpha_{0}\rvert}\right)^{a} \frac{\Gamma[(a+b)/2] \, {}_{1}F_{1}\big((a+b)/2;b+1;(4p)^{-1}\big)}{2^{b+1} p^{(a+b)/2}\Gamma\left(b+1\right)} , \nonumber\\
\end{eqnarray}

\noindent where $G(x)$ is the gamma function, ${}_1F_1(x;y;z)$ is the confluent hypergeometric function, and 
\begin{equation}
a = k+l+2, \quad b =\frac{ k-l}{2}, \quad p = \frac{\overline{n}\left(\overline{n}+1\right)}{4\lvert\alpha_{0}\rvert^2}.
\end{equation}

Using the property $\Gamma(n+1)=n!$ for nonnegative integers $n$ and combining~\eqref{L-general}, \eqref{P-displaced-thermal-app-b}, and \eqref{AUX_INTEGRATION_2}, we finally get 
\begin{eqnarray} \label{L-displaced-thermal_new}
&& \!\!\!\!\! L_{\rm dis.th.}(\theta) = \frac{e^{-|\alpha_0|^2/\overline{n}}}{4\pi \overline{n}} \sum_{\scriptsize \begin{array}{c}
k,l = 0,1,2,\ldots  \\
k+l \text{~is~even}
\end{array}} \frac{\Gamma \left( \frac{3k+l+4}{4} \right)}{\left(\frac{k+l}{2} \right)!\left(\frac{k-l}{2} \right)!} \left( \tfrac{|\alpha_0|^2 \overline{n}}{\overline{n}+1}  \right)^{\frac{k-l}{4}}  \left( \tfrac{\overline{n}}{\overline{n}+1} \right)^{\frac{k+l+2}{2}}  {}_1F_1 \left( \tfrac{3k+l+4}{4}; \tfrac{k-l+2}{2}; \tfrac{|\alpha_0|^2}{\overline{n}(\overline{n}+1)} \right) \nonumber\\ 
&& \!\!\!\!\! \times \exp \left\{ -i [\theta - \theta_0 + g_K t_K (k+l-1)] \frac{k-l}{2} \right\}.
\end{eqnarray}
\twocolumngrid

\end{document}